\documentclass[reprint,amsmath,amssymb,aps,]{revtex4-1}

\usepackage{graphicx}% Include figure files
\usepackage{dcolumn}% Align table columns on decimal point
\usepackage{bm}% bold math
\usepackage{epstopdf}

\usepackage{xcolor}

\begin{document}
\title{Ro-vibrational quenching of C$_2^-$ anions in collisions with He, Ne and Ar atoms}

\author{Barry P. Mant}
 \affiliation{Institut f\"{u}r Ionenphysik und Angewandte Physik, 
Universit\"{a}t Innsbruck, Technikerstr. 25, A-6020, Innsbruck, Austria}
\author{Franco A. Gianturco}
\affiliation{Institut fuer Ionenphysik und Angewandte Physik, 
Universitaet Innsbruck, Technikerstr. 25, A-6020, Innsbruck, Austria}
\email{francesco.gianturco@uibk.ac.at}
\author{Ersin Yurtsever}
\affiliation{Department of Chemistry, Ko\c{c} University, Rumelifeneri yolu, Sariyer, TR-34450, Istanbul, Turkey}
\author{Lola Gonz\'{a}lez-S\'{a}nchez}
\affiliation{Departamento de Qu\'{i}mica F\'{i}sica, University of Salamanca, Plaza de los Ca\'{i}dos sn, 37008 Salamanca, Spain}
\author{Roland Wester}
\affiliation{Institut fuer Ionenphysik und Angewandte Physik, 
Universitaet Innsbruck, Technikerstr. 25, A-6020, Innsbruck, Austria}

\date{\today}% It is always \today, today,

\begin{abstract}
The molecular anion C$_2^-$ is currently of interest as a candidate for laser cooling due to its electronic structure and
favourable branching ratios to the ground electronic and vibrational state. Helium has been proposed as a buffer gas to cool 
the molecule's internal motion. We calculate the cross sections and corresponding rates for ro-vibrational inelastic collisions of
C$_2^-$ with He, and also with Ne and Ar, on new 3D \textit{ab initio} potential energy surfaces using quantum scattering theory. 
The rates for vibrational quenching with He and Ne are very small and are similar to those for small neutral molecules in collision
with helium. 
The quenching rates for Ar however are far larger than those with the other noble gases, suggesting that this may be a more suitable gas 
for driving vibrational quenching in traps. The implications of these new results for laser cooling of C$_2^-$ are discussed.
\end{abstract}

\maketitle

\section{Introduction}
\label{sec:intro}

Laser cooling of molecules has become a very active research area \cite{Tarbutt2018:cp}. With the direct preparation of ultracold molecular
ensembles in magneto-optical traps \cite{Barry2014:n} numerous experiments on molecular quantum control, novel quantum phases \cite{16WoWaHe},
precision spectroscopy \cite{13LoCoGr}, or ultracold chemistry \cite{19DoEbKo} become accessible. For atoms, laser cooling of 
neutral and charged species has developed hand in hand.
For molecules, however, no charged molecular species has yet been successfully laser cooled. Ions with bound excited electronic states that 
lie below the first fragmentation threshold and can be excited with suitable narrow-band lasers are rare. Furthermore, a near optimal 
Franck-Condon overlap of the vibrational wavefunctions is required to make closed optical cycles feasible.

\begin{figure}[htb!]
\centering
\includegraphics[scale=0.33,angle=-90,origin=c]{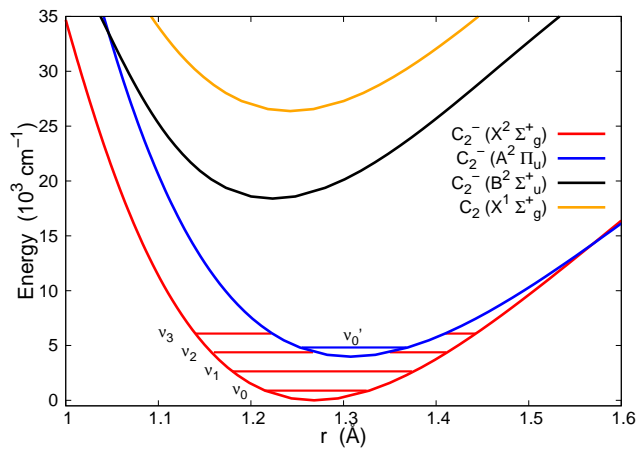}
\vspace*{-15mm}
\caption{Potential energy curves for the three lowest energy electronic states of C$_2^-$ and for the ground state of C$_2$. The vibrational
levels of interest in this study are also shown. The curves were obtained using the RKR method \cite{17RKR1} with the  spectroscopic 
constants from Ervin and Lineberger
 \cite{91ErLixx.c2m}. }
\label{fig:C2m_PECs}
\end{figure}

The diatomic carbon molecular anion has been identified as an interesting exception \cite{15YzHaGe}, as it possesses several bound excited
electronic states below the photodetachment threshold. Furthermore, the 
electronic states $A^2\Pi_u$ and $B^2\Sigma_u^+$ (Fig.  \ref{fig:C2m_PECs}) have high Franck-Condon overlap factors with 
the $X ^2\Sigma^+_g$ ground state for the transitions between their lowest vibrational levels $\nu'=0 \rightarrow \nu''=0$
 \cite{03ShXiBe,16ShLiMe.c2m}. Simulations of laser cooling
using the $B^2\Sigma_u^+$ \cite{15YzHaGe} and $A^2\Pi_u$ \cite{17FeGeDo} states have both shown that C$_2^-$ can, in principle, be cooled
efficiently to milikelvin temperatures using Doppler or Sisyphus cooling in Paul or Penning traps. Photodetachment cooling has also been 
shown to allow even lower temperatures to be accessed \cite{18GeFeDo}. If laser cooling of C$_2^-$ anions were to be realised, 
it would open up the possibility of sympathetically cooling other anions \cite{17FeGeDo} or even antiprotons \cite{18GeFeDo}. 
This last achievement could allow the efficient
production of antihydrogen atoms, currently being investigated for tests of fundamental physics such as CPT invariance \cite{17Ahxxxx} 
and the weak equivalence principle \cite{12PeSaxx}.

The diatomic carbon molecular anion C$_2^-$ has been a model system for decades, attracting a great deal of experimental 
\cite{68HeLaxx.c2m,69MiMaxx.c2m,71.Frxxxx.c2m,72LiPaxx.c2m,80JoMeKo.c2m,82LeNaMi.c2m,85MeHeSc.c2m,88ReLiDi.c2m,91ErLixx.c2m,
92RoZaxx.c2m,95BeZhYu.c2m,98PeBrAn.c2m,03BrWeDa.c2m,17Naxxxx.c2m,14EnLaHa.c2m} and theoretical 
\cite{74Baxxxx.c2m,79ZePeBu.c2m,80DuLixx.c2m,84RoWexx.c2m,87NiSixx.c2m,92WaBaxx.c2m,06SeSpxx.c2m,16ShLiMe.c2m,19KaLoLi.c2m,19GuJaKr} work. 
Its bound electronically excited states \cite{87NiSixx.c2m} are unusual for an anion, which is a consequence of the high electron affinity 
of neutral C$_2$ of around 3.3 eV \cite{80JoMeKo.c2m,91ErLixx.c2m} in combination with the open shell character of the electronic
configuration of carbon dimers. In its ground electronic state $X ^2\Sigma^+_g$ the molecule has only evenly numbered rotational states due
to the nuclear statistics of the $^{12}$C$_2^-$ molecule with zero spin nuclei, while in the excited $B^2\Sigma_u^+$ state only odd-numbered
rotational states exist.

It has also been suggested that C$_2^-$ could be present in astronomical environments as neutral C$_2$ is abundant in interstellar space
\cite{95LaShFe.c2m}, comet tails \cite{90LaShDa.c2m} and is a common component of 
carbon stars \cite{77SoLuxx.c2m,86LaGuEr.c2m}. The large EA of C$_2$ and strong electronic absorption bands of 
C$_2^-$ \cite{72LiPaxx.c2m} suggest that the anion could also be detected in space \cite{80VaSwxx.c2m} but as yet no conclusive evidence of
its presence has been found \cite{72FaJoxx.c2m,82Waxxxx.c2m,05CiHoKa.c2m}. As the most abundant isotopologue $^{12}$C$_2^-$
is a homonuclear diatomic molecule, it does not exhibit a pure ro-vibrational spectrum making its detection in emission difficult.
Transitions to and from low lying excited electronic states could however allow for the anions detection or, as will be evaluated and
discussed here, as a possible option, for the detection of the $^{12}$C$^{13}$C$^-$ isotopologue which would then have a small dipole moment.

Laser cooling of C$_2^-$ would ideally start with ions initially cooled to around 10 K, for example by helium buffer gas cooling in a
cryogenic ion trap \cite{Wester2009:jpb,19GiGoMa.c2hm}. Processes used to generate C$_2^-$ involve applying an electric discharge to a 
mixture of C$_2$H$_2$ and CO$_2$ in a carrier gas \cite{03BrWeDa.c2m,19HiGeOs} which may form the anion in excited vibrational states.
Besides cooling the translational motion, the buffer gas is then also required to cool internal degrees of freedom via inelastic collisions.
Furthermore, buffer gas may be a useful tool to quench excited vibrational levels when they get populated during laser cooling due to
the non-diagonal Franck-Condon factors. This could circumvent the need for additional repumping lasers. In a similar scheme, 
rotational buffer gas cooling was performed during sympathetic translational cooling of MgH$^+$ \cite{14HaVeKl}.

In a recent paper we calculated cross sections and rate coefficients for C$_2^-$-He rotationally inelastic collisions, treating the anion 
as a rigid rotor \cite{20MaGiGo}. The rates for rotational excitation and quenching were found to be in line with those 
for similar ionic molecules interacting with helium \cite{19GiGoMa.c2hm}. Simulations of cooling rotational motion at typical helium 
pressures in ion traps showed thermalisation to Boltzmann populations occurred within tenths of seconds. Very recently we have extended 
this work and also modelled the rotational cooling of C$_2^-$ with neon and argon \cite{20aMaGiWe}. It was found that thermalization 
times of C$_2^-$ with He and Ne were fairly similar but cooling was significantly faster with Ar. This is due to the increased interaction
strength between C$_2^-$ and the larger atoms which increased as expected in the series He $<$ Ne $<$ Ar. 

In this work we present results for the quenching of internal vibrational motion of C$_2^-$ in its ground $^2 \Sigma_g^+$ electronic state
in collisions with the noble gas atoms helium, neon and argon. A simplified view of the relevant vibrational levels involved in these
processes is shown specifically in Figure  \ref{fig:C2m_PECs}. We know, however, that no quantitative rate coefficients for vibrational
relaxation are available to date. As C$_2^-$ has no oscillating dipole, the vibrational levels are long lived with the ground electronic
state's $v=2$ levels persisting for over five seconds \cite{98PeBrAn.c2m} and so collisions are the only viable means of quenching these
states efficiently. The rate coefficients of C$_2^-$ vibrational quenching with helium may also prove useful in future astronomical studies, 
should the anion be detected in an interstellar environment where excited vibrational states are important for observation, as in the
circumstellar envelope around carbon rich stars where helium atoms are also abundant.

The paper is organised as follows. In the next section we discuss the potential energy curve (PEC) and vibrational levels of the
isolated C$_2^-$ in its ground electronic $^2 \Sigma_g^+$ state. We further calculate the dipole moment of the $^{12}$C$^{13}$C$^-$
isotopologue and discuss its value. In Section
\ref{sec:3DPES} we provide details of \textit{ab initio} calculations for the 3D potential energy surfaces (PES) and fitting of the
surfaces to a functional form. This section also contains details of the vibrationally averaged matrix elements required for 
scattering calculations. Details of the close-coupled scattering calculations are given in Section \ref{sec:scat}. Cross sections 
and corresponding rates for rotationally and vibrationally inelastic collisions are presented in Section \ref{sec:results}. We present
conclusions in Section \ref{sec:conc}.

\section{C$_2^-$ ($^2 \Sigma_g^+$) potential energy curve and $^{12}$C$^{13}$C$^-$ dipole moment} 
\label{sec:PEC}

As recently discussed by Gulania \textit{et al.} \cite{19GuJaKr}, the electronic structure of the C$_2$ molecule is notoriously 
difficult to calculate accurately due to many low-lying electronic states giving rise to a multireference character of the  
\textit{ab initio} description of its ground electronic state. For the C$_2^-$ anion considered here, the 
situation is not so severe but the presence of a close-lying $A^2\Pi_u$ state (4000 cm$^{-1}$ above the ground $^2 \Sigma_g^+$ state, see Fig.
 \ref{fig:C2m_PECs}) still makes electronic structure calculations challenging. 

The PEC of C$_2^-$ in its ground $^2 \Sigma_g^+$ state was calculated from the 3D potential energy surfaces (see next section)  
with the noble gas atom at $R = 25$ \AA. 
The LEVEL program \cite{17LEVEL} was used to obtain the
vibrational energies and wavefunctions for the C$_2^-$ molecule. \textit{Ab initio} PEC points were used as input, interpolated
using a cubic spline and extrapolated to $r$ values below and above our range using functions implemented in LEVEL. 
The relative energies of the first three vibrational levels along with the rotational constants for each state are shown in 
Table \ref{tab:VibE} and compared with previously published calculated theoretical and experiment values. Table \ref{tab:VibE} also 
compares the values obtained for the MCSCF method which was used for the C$_2^-$-He PES and the CCSD-T method which was used for 
C$_2^-$-Ne/Ar (see next section). While we do not achieve 
spectroscopic accuracy with our PEC, the relative energy spacings are sufficiently realistic for computing the vibrational quenching rates of
interest here at a reliable level. The results obtained for CCSD-T are closer in agreement to experiment than those for the 
MCSCF method but the differences will have a minimal impact on the computed inelastic rate coefficients obtained from our scattering
calculations of interest here. The PEC fit using LEVEL to the CCSD-T calculations and 
vibrational wavefunctions for $\nu= 1,2$ and 3 are provided in the Supplementary Material \cite{pra_supp}.

\begin{table}[h!]
\setlength{\tabcolsep}{15pt}
\renewcommand{\arraystretch}{1.2}
\caption{\label{tab:VibE} Comparison of vibrational energies and rotational constants with previous theoretical and experimental values.
Literature values calculated from Dunham parameters provided. Units of cm$^{-1}$.}
\begin{tabular}{cccc}
\hline                        
     &   & Relative energy  & $B_{\nu}$    \\
\hline
 $\nu_0$ & MCSCF & 0  & 1.7455  \\
         & CCSD-T & 0  & 1.7356  \\
         & Calc. \cite{16ShLiMe.c2m} & 0 & 1.7358 \\
         & Exp. \cite{85MeHeSc.c2m} & 0 & 1.7384 \\
 $\nu_1$ & MCSCF & 1805  & 1.7419  \\
         & CCSD-T & 1776  & 1.7222  \\
         & Calc. \cite{16ShLiMe.c2m} & 1759 & 1.7197 \\
         & Exp. \cite{85MeHeSc.c2m} & 1757 & 1.7220 \\
 $\nu_2$ & MCSCF & 3633  & 1.7190  \\
         & CCSD-T & 3561  &  1.7126  \\
         & Calc. \cite{16ShLiMe.c2m} & 3494 & 1.7035 \\
         & Exp. \cite{85MeHeSc.c2m} & 3492 & 1.7062 \\
\hline
\end{tabular}
\end{table}

As discussed above, the C$_2$ molecule has been detected in various astronomical settings \cite{95LaShFe.c2m,90LaShDa.c2m,
77SoLuxx.c2m,86LaGuEr.c2m} but searches for the C$_2^-$ anion focusing on electronic transitions have so far not been conclusive
\cite{72FaJoxx.c2m,82Waxxxx.c2m,05CiHoKa.c2m}. Franck-Condon factors and Einstein A coefficients for these transitions have been
calculated by Shi \textit{et al.} \cite{16ShLiMe.c2m}. Another possible detection method, at least in principle, is the rotational
transitions of the $^{13}$C$^{12}$C$^-$ isotopologue \cite{06SeSpxx.c2m}. The rotational constants for this isotoplogue were 
accurately calculated by \u{S}edivcov\'{a} and \u{S}pirko \cite{06SeSpxx.c2m}. Here we use our PEC to assess the dipole moment of
$^{13}$C$^{12}$C$^-$ and Einstein A coefficients for rotational transitions. 

The dipole moment of a charged homonuclear diatomic with different isotopes arises due to the difference in the centre of mass and
centre of charge. An expression for the dipole moment of HD$^+$ was derived by Bunker \cite{74Buxxxx} and Ellison \cite{62Elxxxx} as
\begin{equation}
\mu(\nu',\nu) = - [(m_a - m_b)/2m_T] e \langle \nu' |r| \nu \rangle
\label{eq.dipole}
\end{equation}
where $m_a$ and $m_b$ are the masses of each nucleus and $m_T = m_a + m_b$. Using LEVEL the matrix element of the vibrational coordinate
$r$ for $^{13}$C$^{12}$C$^-$ for the $\nu' = \nu = 0$ ground vibrational state was calculated as 1.27 \AA, close to the equilibrium
geometry of C$_2^-$ of $r_{eq} = 1.2689$ \AA \cite{16ShLiMe.c2m}. Using the masses for $^{13}$C$^{12}$C$^-$ in Eq. \ref{eq.dipole} gives 
$\mu(0,0) = 0.12$ D. This compares to 0.87 D in HD$^+$ \cite{74Buxxxx}. For pure rotational transitions the Einstein coefficient for
spontaneous dipole transitions is given as \cite{19GiGoMa.c2hm}
\begin{equation}
A_{k \rightarrow i} = \frac{2}{3} \frac{\omega_{k \rightarrow i}^3}{\epsilon_0 c^3 h} \mu_0^2 \frac{j_k}{(2 j_k + 1)}
\label{eq:dip_tran_rot}
\end{equation}
where $\omega_{i \to k} \approx 2B_0(j_i + 1)$ is the transition's angular frequency. In Table \ref{tab:EinA} the Einstein A coefficients
computed using Eq. \ref{eq:dip_tran_rot} for $^{12}$C$^{13}$C$^-$ (treated as pseudo-singlet), HD$^+$ and C$_2$H$^-$ are compared 
for the first few rotational levels. The Einstein A coefficients for $^{12}$C$^{13}$C$^-$ are orders of magnitude smaller than for HD$^+$ 
and C$_2$H$^-$ and other molecular ions \cite{19GiGoMa.c2hm}. The combination of very small rotational emission coefficients coupled with 
the isotope ratio for $^{13}$C/$^{12}$C of 0.01 suggests that detecting C$_2^-$ via the rotational transitions of the $^{12}$C$^{13}$C$^-$
isotopologue would be very difficult.

\begin{table}[h!]
\setlength{\tabcolsep}{10pt}
\renewcommand{\arraystretch}{1.2}
\caption{\label{tab:EinA} Computed Einstein spontaneous emission coefficients $A_{j \to j'}$ for $^{12}$C$^{13}$C$^-$ ($B_0$ = 
1.67152 cm$^{-1}$ \cite{06SeSpxx.c2m}, $\mu = 0.12$ D), HD$^+$ ($B_e$ = 22.5 cm$^{-1}$ \cite{12IsNaNa}), $\mu$ = 0.87 D \cite{74Buxxxx}
and C$_2$H$^-$ ($B_e$ = 1.389 cm$^{-1}$ \cite{12DuSpSe}, $\mu$ = 3.09 D \cite{07BrGoGu} ). 
All quantities in units of s$^{-1}$. } 
\begin{tabular}{cccc}
\hline                           
     Transition   & $^{12}$C$^{13}$C$^-$ & HD$^+$  & C$_2$H$^-$   \\ 
\hline  
$1 \to 0$ & 5.6$\times 10^{-8}$ & 7.2$\times 10^{-3}$ & 2.14$\times 10^{-5}$ \\
$2 \to 1$ & 5.4$\times 10^{-7}$ & 6.9$\times 10^{-2}$ & 2.05$\times 10^{-4}$ \\
$3 \to 2$ & 1.9$\times 10^{-6}$ & 2.5$\times 10^{-1}$ & 7.43$\times 10^{-4}$ \\
$4 \to 3$ & 4.8$\times 10^{-6}$ & 6.0$\times 10^{-1}$ & 1.83$\times 10^{-3}$ \\
$5 \to 4$ & 9.6$\times 10^{-6}$ & 1.2$\times 10^{0}$  & 3.65$\times 10^{-3}$ \\
\hline
\end{tabular}
\end{table}

\section{C$_2^-$-H\MakeLowercase{e}/N\MakeLowercase{e}/A\MakeLowercase{r} 3D potential energy surfaces and vibrationally 
averaged matrix elements}
\label{sec:3DPES}

\begin{figure*}[htb!]
\centering
\includegraphics[scale=0.5,angle=-90,origin=c]{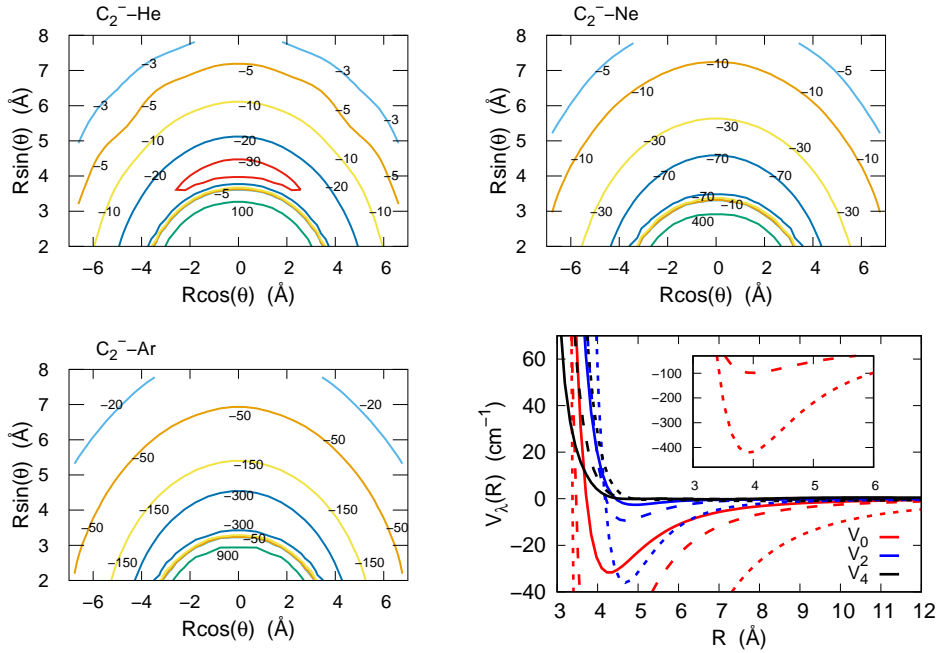}
\vspace*{-23mm}
\caption{Contour plots of C$_2^-$($^2 \Sigma_g^+$)-He (top left), Ne (top right) and Ar (bottom left) 
vibrationally averaged matrix elements $V_{0,0}(R,\theta)$ projected onto Cartesian coordinates. Energies in cm$^{-1}$. Bottom right is 
expansion of matrix elements in V$_{\lambda}$ coefficients for $V_{0,0}$ for He (solid lines), Ne (long dashed lines) and Ar 
(short dashed lines). V$_0$ in red (light grey), V$_2$ in blue (darker grey) and V$_4$ in black. }
\label{fig:V00}
\end{figure*}

The interaction energies between C$_2^-$ in its ground $^2 \Sigma_g^+$ electronic state with He, Ne and Ar atoms were calculated 
using \textit{ab initio} methods implemented in the MOLPRO suite of codes  \cite{MOLPRO,MOLPRO_brief}. Geometries were defined on a 
Jacobi grid with $R$ (the distance from the centre of mass of C$_2^-$ to the atom) ranging from 2.6 to 25 {\AA} and $\theta$ (the angle
between $R$ and the C$_2^-$ internuclear axis $r$) from 0 to 90$^{\circ}$ in 10$^{\circ}$ intervals. Five values of the C-C bond length for
each system between $r=1.10$-1.35 \AA $ $ were used including the equilibrium value of $r_{eq} = 1.269$ \AA. This is sufficient to cover 
the vibrational levels of interest in the present study. Interaction potential energies between C$_2^-$ and the noble gas atoms were
determined by subtracting the asymptotic energies for each bond length. 

For C$_2^-$-He, energies were calculated using the 2x2 Multi-configurational self-consistent 
field (MCSCF) method \cite{85WeKnxx,85KnWexx} with 10 occupied orbitals and 4 closed orbitals followed by a 2-state multi-reference
configuration interaction (MRCI) \cite{11ShKnWe.LM} calculation. An aug-cc-pVQZ basis \cite{92KeDuHa} was used on each carbon centre and 
an aug-cc-pV5Z basis on the helium atom. For the C$_2^-$-Ne and Ar systems, convergence problems were encountered for the MCSCF approach 
and so energies were instead calculated using the RCCSD-T method for open shell systems \cite{93KnHaWe,94DeKnxx} 
with complete basis set (CBS) extrapolation using 
the aug-cc-pVTZ, aug-cc-pVQZ and aug-cc-pV5Z basis sets \cite{96WiMoDu,93WoDuxx}. The same method for the case of C$_2^-$-He provided results
within a few wavenumbers of the MCSCF approach. The basis-set-superposition-error (BSSE) was also accounted for at all calculated points 
using the counterpoise procedure \cite{70BoBexx}.  

The three-dimensional PESs were fit to an analytical form using the method of Werner, Follmeg and Alexander \cite{88WeFoAl,
17BaDaxx} where the interaction energy is given as
\begin{equation}
V_{\mathrm{int}}(R,r,\theta) = \sum_{n=0}^{N_r -1} \sum_{l=0}^{N_{\theta}-1} P_l (\cos \theta) A_{ln}(R) (r-r_{eq})^n,
\label{eq.PES_func}
\end{equation}
where $N_r$ = 5 and $N_{\theta}$ = 10 are the number of bond lengths $r$ and angles $\theta$ in our \textit{ab initio} grid, 
$P_l (\cos \theta)$ are the Legendre polynomials where due to the symmetry around $\theta = 90^{o}$ only even values of $l$ are used 
and $r_{eq} = 1.2689$ {\AA}   is the equilibrium bond length of C$_2^-$. For each bond length $r_m$ and angle $\theta_k$, one-dimensional 
cuts of the PESs $V_{\mathrm{int}}(R,r_m,\theta_k)$ were fit to
\begin{eqnarray}
B_{km}(R) = \exp(-a_{km}R) \left[ \sum_{i=0}^{i_{\mathrm{max}}} b_{km}^{(i)} R^i \right] \nonumber \\
- \frac{1}{2} \left[1 + \tanh(R) \right] 
\left[ \sum_{j=j_{\mathrm{min}}}^{j=j_{\mathrm{max}}} c_{km}^{j} R^{-j} \right],
\label{eq.rad_fit}
\end{eqnarray}
where the first terms account for the short range part of the potential and the second part for the long range terms combined using
the $\frac{1}{2} \left[1 + \tanh(R) \right]$ switching function. For each $r_m$ and $\theta_k$ Eq. \ref{eq.rad_fit} was 
least squares fit to the \textit{ab initio} data (around 40 $R$ points) using $i_{\mathrm{max}} = 2$, $j_{\mathrm{min}} = 4$ and
$j_{\mathrm{max}} = 10$ for eight variable parameters. The average root-mean-square error (RMSE) for each fit was 0.5 cm$^{-1}$ for 
C$_2^-$-He and Ne, rising to 1 cm$^{-1}$ for Ar. From the 1D potential fits $B_{km}(R)$, the radial coefficients $A_{ln}(R)$ can be 
determined from the matrix product $\mathbf{A}(R) = \mathbf{P}^{-1}\mathbf{B}(R)\mathbf{S}^{-1}$ where the matrix elements of 
$\mathbf{P}$ and $\mathbf{S}$ are given as
$P_{kl} = P_l(\cos \theta_k)$ and $S_{nm} = (r_m - r_{eq})^n$ respectively. The analytical representation of the PES, Eq.
\ref{eq.PES_func}, gives a reasonable representation of the \textit{ab initio} interaction energies. An RMSE of 
1.5 cm$^{-1}$ for $V < 200$ cm$^{-1}$ and 0.9 cm$^{-1}$ for $V < 0$ cm$^{-1}$ was obtained for the C$_2^-$-He system while for 
the C$_2^-$-Ne and Ar systems RMSEs of 0.5 and 3.5 cm$^{-1}$ respectively for $V < 1500$ cm$^{-1}$ were obtained.  

The scattering calculations described in the next section require the interaction potential to be averaged over the vibrational states
of C$_2^-$ as
\begin{equation}
V_{\nu,\nu'}(R,\theta) = \langle \chi_{\nu}(r) | V_{\mathrm{int}}(R,r,\theta) | \chi_{\nu'}(r) \rangle.
\label{eq.vib_coup}
\end{equation}
Fig. \ref{fig:V00} shows the diagonal terms $V_{0,0}(R,\theta)$ for each system. As expected for a molecule with a strong bond, the contour
plots of the $V_{0,0}(R,\theta)$ for each system are very similar to our earlier rigid-rotor (RR) PESs  which were obtained
without the vibrational averaging (and a different \textit{ab initio} method for C$_2^-$-He)  \cite{20MaGiGo,20aMaGiWe}. 
The minimum values of $V_{0,0}$ for each system occur at perpendicular 
geometries and are around $-30$ cm$^{-1}$ at 4.5 {\AA} for He, $-110$ cm$^{-1}$ at 3.7 {\AA} for Ne and $-490$ cm$^{-1}$ at 3.7 {\AA} for Ar.
Each system's PES has a fairly similar appearance with the well depth being the main
difference which increases as expected from He to Ne to Ar due to the increasing number of electrons on the
atoms and on the much larger dipole polarizabiliy that dominates the long-range attractive terms with a value of 1.383 $a_0^3$ for He,
2.660  $a_0^3$ for Ne, and 11.070 $a_0^3$ for Ar \cite{18GaFexx}.

\begin{figure*}[htb!]
\centering
\includegraphics[scale=0.5,angle=-90,origin=c]{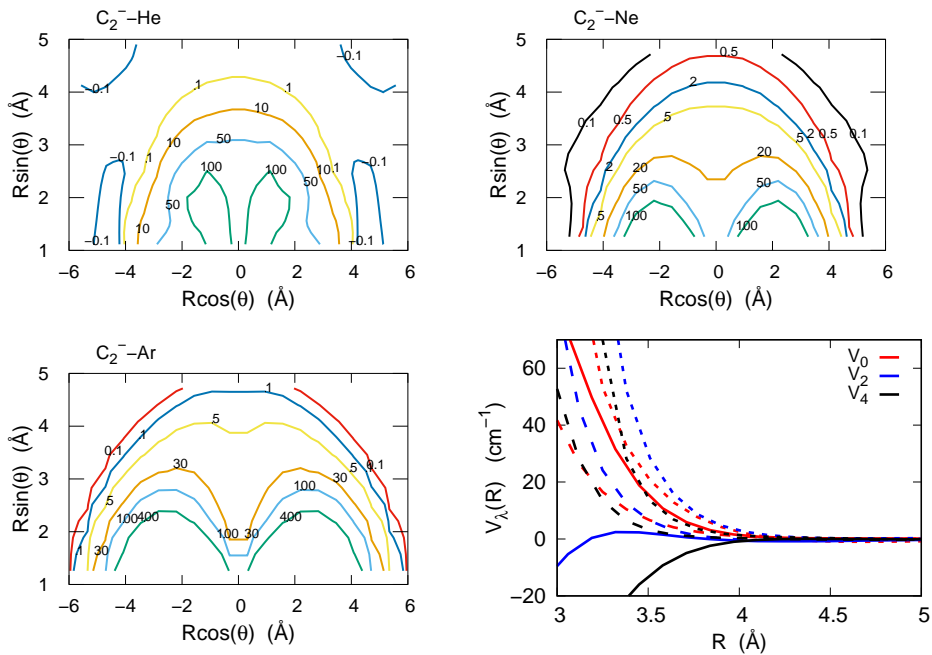}
\vspace*{-23mm}
\caption{Contour plots of C$_2^-$($^2 \Sigma_g^+$)-He (top left), Ne (top right) and Ar (bottom left) 
vibrationally averaged matrix elements $V_{0,1}(R,\theta)$ projected onto Cartesian coordinates. Energies in cm$^{-1}$. Bottom right is 
expansion of matrix elements in V$_{\lambda}$ coefficients for $V_{0,1}$ for He (solid lines), Ne (long dashed lines) and Ar 
(short dashed lines). V$_0$ in red (light grey), V$_2$ in blue (darker grey) and V$_4$ in black.}
\label{fig:V01}
\end{figure*}

The off diagonal $V_{0,1}(R,\theta)$ terms which directly drive vibrationally inelastic $\nu = 1$ to $\nu=0$ transitions are shown in 
Fig. \ref{fig:V01}. At short distances the coupling terms are repulsive, becoming negligible quickly at longer distances, as is the 
case for many other atom-diatom systems. It can be seen that for C$_2^-$ interacting with He and Ne the $V_{0,1}(R,\theta)$ plots are quite
similar but the interaction with Ar is more repulsive. This suggests collisions with Ar have larger vibrational cross sections as will be 
shown below. The $V_{0,2}(R,\theta)$ and $V_{1,2}(R,\theta)$ matrix elements have a similar appearance to those of $V_{0,1}(R,\theta)$.

The close-coupling scattering calculations discussed the next section require the vibrationally averaged matrix elements in the
form of a multipole expansion as
\begin{equation}
V_{\nu,\nu'}(R,\theta) = \sum_{\lambda}^{\lambda_{\rm{max}}} V_{\nu,\nu'}^{\lambda}(R) P_{\lambda} (\cos \theta)
\label{eq.Legendre}
\end{equation}
where again due C$_2^-$ being a homonuclear diatomic, only even $\lambda$ terms are required. The bottom right panels of Figures 
\ref{fig:V00} and \ref{fig:V01} compare the $V_{0,0/1}^{\lambda}(R)$ coefficients for the most important $\lambda = 0, 2$ and 4 coefficients.
As with the contour plots of Figure \ref{fig:V00}, the $V_{0,0}^{\lambda}(R)$ expansion coefficients are very similar to their rigid-rotor
counterparts \cite{20MaGiGo,20aMaGiWe}. This means that rotationally inelastic collisions using the vibrationally averaged multipole
expansion will have very similar values to those obtained using a rigid-rotor treatment as will be shown in Section \ref{sec:rot_xsec}. 
As for other atom-diatom systems, $V_{\nu,\nu}^{\lambda}(R)$ for other vibrational states $\nu$ are very similar to those for
$V_{0,0}^{\lambda}(R)$ and thus rotations and vibrations can essentially be considered separately.

For the off-diagonal expansion coefficients $V_{0,1}^{\lambda}$, all terms quickly approach zero as $R$ is
increased. For all three systems the  $V_{0,1}^{0}(R)$ coefficients are steeply repulsive as $R$ decreases. For the C$_2^-$-He system 
however the $V_{0,1}^{2}(R)$ and $V_{0,1}^{4}(R)$ terms are attractive in contrast to Ne and Ar which are also repulsive. As expected 
from the contour plots, the $V_{0,1}^{\lambda}(R)$ terms are the most repulsive for the C$_2^-$-Ar interaction.

The PES functions and vibrationally averaged matrix elements  used for each system 
are provided in the Supplementary Material \cite{pra_supp}.

\section{Quantum scattering calculations} 
\label{sec:scat}

Quantum scattering calculations were carried out using the coupled channel (CC) method to solve the Sch\"{o}dinger equation for
scattering of an atom with a diatomic molecule as implemented in our in-house code, ASPIN \cite{08LoBoGi}.
The method has been described in detail before \cite{60ArDaxx,08LoBoGi} 
and only a brief summary will be given here, with equations given in atomic units. 
For a given total angular momentum $\mathbf{J = l + j}$ the scattering 
wavefunction is expanded as 
\begin{equation}
\Psi^{JM}(R, r,  \Theta) = \frac{1}{R} \sum_{\nu,j,l} f_{\nu lj}^J (R) \chi_{\nu,j}(r) \mathcal{Y}_{jl}^{JM}(\hat{\mathbf{R}},
\hat{\mathbf{r}}),
\label{eq.basis}
\end{equation}
where $l$ and $j$ are the orbital and rotational angular momentum respectively, $\mathcal{Y}_{jl}^{JM}(\hat{\mathbf{R}},
\hat{\mathbf{r}})$ are coupled-spherical harmonics for $l$ and $j$ which are eigenfunctions of $J$. $\chi_{\nu,j}(r)$ are the 
radial part of the ro-vibrational eigenfunctions of the molecule.
The values of $l$ and $j$ are constrained, via Clebsch-Gordan coefficients, such that their resultant summation is 
compatible with the total angular momentum $J$ \cite{60ArDaxx,08LoBoGi}.
$f_{\nu lj}^J (R)$ are 
the radial expansion functions which need to be determined. Substituting the expansion into the Sch\"{o}dinger equation 
with the Hamiltonian for atom-diatom scattering \cite{60ArDaxx,08LoBoGi} leads to the  CC equations for each $J$
\begin{equation}
\left(\frac{d^2}{dR^2} + \mathbf{K}^2 - \mathbf{V} - \frac{\mathbf{l}^2}{R^2} \right) \mathbf{f}^J = 0.
\label{eq.CC}
\end{equation}
Here each element of $\mathbf{K} = \delta_{i,j}2 \mu (E- \epsilon_i)$ (where $\epsilon_i$ is the channel asymptotic energy),
$\mu$ is the reduced mass of the system, $\mathbf{V}= 2 \mu \mathbf{U}$ is the interaction potential matrix between
channels and $\mathbf{l}^2$ is the matrix of orbital angular momentum. For the ro-vibrational scattering calculations of interest here,
the matrix elements $\mathbf{U}$ are given explicitly as
\begin{eqnarray}
\langle \nu j l J | V | \nu' j' l' J \rangle = \int_0^{\infty} \mathrm{d}r \int \mathrm{d} \hat{\mathbf{r}} \int \mathrm{d} \hat{\mathbf{R}} 
\nonumber \\
\chi_{\nu,j} (r) \mathcal{Y}_{jl}^{JM}(\hat{\mathbf{R}},\hat{\mathbf{r}})^* |V(R,r,\theta)| \chi_{\nu',j'}(r) \mathcal{Y}_{j'l'}^{JM}
(\hat{\mathbf{R}},\hat{\mathbf{r}}).
\label{eq.vib_elements}
\end{eqnarray}
As the intermolecular potential $V(R,r,\theta)$ is expressed as in Eq. \ref{eq.Legendre}, Eq. \ref{eq.vib_elements} can be written as
\begin{equation}
\langle \nu j l J | V | \nu' j' l' J \rangle = \sum_{\lambda=0}^{\infty} V_{\nu,\nu'}^{\lambda}(R) f^J_{\lambda j l j' l'},
\label{eq.vib_elements2}
\end{equation}
where the $f^J_{\lambda j l j' l'}$ terms are the Percival-Seaton coefficients
\begin{equation}
f^J_{\lambda j l j' l'} = \int \mathrm{d} \hat{\mathbf{r}} \int \mathrm{d} \hat{\mathbf{R}} \quad \mathcal{Y}_{jl}^{JM}(\hat{\mathbf{R}}
,\hat{\mathbf{r}})^* P_{\lambda}(\cos \theta) \mathcal{Y}_{j'l'}^{JM}(\hat{\mathbf{R}},\hat{\mathbf{r}}),
\label{eq.Percival-Seaton}
\end{equation}
for which analytical forms are known \cite{08LoBoGi}. Eq. \ref{eq.vib_elements2} also makes use of the widely known approximation
\begin{equation}
V_{\nu,\nu'}^{\lambda}(R) \approx V_{\nu j \nu' j'}^{\lambda}(R),
\label{eq.vib_approx}
\end{equation}
for all $j$ such that the effect of rotation on the vibrational matrix elements is ignored.

The CC equations are propagated outwards from the
classically forbidden region to a sufficient distance where the scattering matrix $\mathbf{S}$ can be obtained. The 
ro-vibrational state-changing cross sections are obtained as
\begin{equation}
\sigma_{\nu j \rightarrow \nu j'} = \frac{\pi}{(2j+1)k_{\nu j}^2} \sum_J (2J+1) \sum_{l,l'} | \delta_{\nu lj, \nu'l'j'} - 
S^J_{\nu lj,\nu' l'j''} |^2.
\end{equation}

In all scattering calculations the C$_2^-$ anion was treated as pseudo-singlet ($^1 \Sigma$) and the effects of 
spin-rotation coupling were ignored. In our previous work on this system it was shown that a pseudo-singlet treatment of the rotational 
state-changing collisions resulted in essentially the same results as the explicit doublet calculation when the relevant cross sections 
were summed \cite{20MaGiGo}. This approximation reduces the computational cost of the scattering calculations without significantly 
affecting the size of the cross sections and thus the main conclusions.

To converge the CC equations, a rotational basis set was used which included up to $j=20$ rotational functions for each vibrational state. 
The CC equations were propagated between 1.7 and 100.0 {\AA} using the log-derivative propagator \cite{86Maxxxx.c2m} up to 
60 {\AA} and the variable-phase method at larger distances \cite{03MaBoGi}. The potential energy was interpolated between calculated
$V_{\nu,\nu'}^{\lambda}(R)$ values using a cubic spline. For $R < 2.6$ \AA $ $ the $V_{\nu,\nu'}^{\lambda}(R)$ were extrapolated
as $\frac{a_{\lambda}}{R} + b_{\lambda}R$ while for $R > 20$ \AA $ $ the $\lambda = 0$ terms were extrapolated as 
$\frac{c}{R^4} + \frac{d}{R^6}$. As our \textit{ab initio} calculated interaction energies were computed to $R = 25$ \AA $ $ where the
interaction energy is negligible for the temperature of interest here, the extrapolated form has also a  negligible effect on 
cross sections \cite{20MaGiGo}.

A number of parameters of the calculation were checked for convergence. The number of $\lambda$ terms from Eq. \ref{eq.Legendre}
was checked for both rotationally and vibrationally inelastic collisions. For the former, calculations were converged to better than 
1 \% using only three terms (up to $\lambda = 4)$. For vibrationally inelastic collisions the convergence with increasing $\lambda$ are less
precise: with five $\lambda$ terms convergence to tens of percent is achieved for He and Ne. For Ar convergence to within about a factor of
two is achieved. This is due to the very small cross sections for these processes which makes obtaining precise and stable values 
more difficult to achieve. For production calculations, nine $\lambda$ terms were included for each $V_{\nu,\nu'}(R)$.

The effect of the PES fitting function was also checked. For the C$_2^-$-He system a PES fit was carried out using only three $r$ 
terms with $r= 1.10, 1.2689$ and 1.35 \AA. This change resulted in tens of percent changes to the vibrationally inelastic cross sections. 
The fitting function, Eq. \ref{eq.PES_func}, does not extrapolate well and this change from a fifth to third order polynomial fit for $r$ 
has a drastic effect on the variation in potential energy with $r$ for values below and above the range used for fitting. 
Despite this, the vibrationally inelastic cross sections remained reasonably consistent and thus our $r$ range is sufficient 
to obtain cross sections which are to the correct order of magnitude, sufficient to assess rates for vibrational 
quenching of C$_2^-$ with each of the noble gas atoms.

As a final check of our calculation parameters, the effect of the vibrational basis set was also considered. In all calculations we used the
vibrational energies and rotational constants obtained from calculations using LEVEL and employing our own C$_2^-$ PEC as discussed in section
\ref{sec:PEC}. It was found that for the $\nu = 1$ and $\nu =2$, which are the  states of interest here (see next section),
it was sufficient to only include these states. Including the $\nu = 3$ state had a negligible effect on the $\nu = 1$ and $\nu =2$ 
quenching cross sections.

Scattering calculations were carried out for collision energies between 1 and 1000 cm$^{-1}$ using steps of 0.1 cm$^{-1}$ for 
energies up to 100 cm$^{-1}$, 0.2 cm$^{-1}$ for 100-200 cm$^{-1}$, 1.0 cm$^{-1}$ for 200-300 cm$^{-1}$, 2 cm$^{-1}$ for
300-700 cm$^{-1}$ and 4 cm$^{-1}$ for 700-1000 cm$^{-1}$. This fine energy grid was used to ensure that important features such as 
resonances appearing in the cross sections were accounted for and their contributions correctly included when the corresponding
rates were calculated. At low collision energies, such resonances will be very sensitive to the details of the PES 
(see below).
The number of partial waves was increased with increasing energy up to $J = 100$ for the highest energies considered.

\section{Results}
\label{sec:results}

\subsection{Rotationally inelastic cross sections}
\label{sec:rot_xsec}

\begin{figure}[h!]
\centering
\includegraphics[scale=0.33,angle=-90,origin=c]{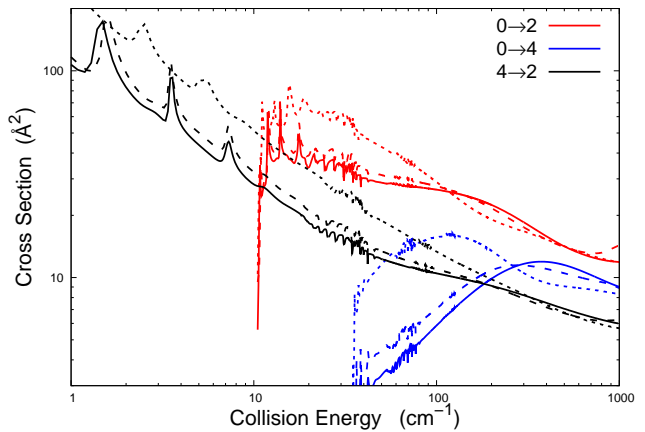}
\vspace*{-15mm}
\caption{Rotationally inelastic cross sections computed using vibrationally averaged method (solid lines), using a RR approach
with the current PES for $r=r_{eq}$ (long-dashed lines) and those of our previous RR PES (short-dashed lines) \cite{20MaGiGo}.}
\label{fig:rot_xsec_comp}
\end{figure}

Rotationally inelastic cross section for C$_2^-$-He collisions can be used to compare our present calculations with those of our
previous work which considered rotationally inelastic collisions treating the anion as a rigid-rotor \cite{20MaGiGo}. Fig.
\ref{fig:rot_xsec_comp} shows rotationally inelastic cross sections for selected $j \rightarrow j'$ transitions for both excitation
and de-excitation processes. The figure compares using a vibrationally averaged (VA) PES, that is, only $V_{0,0}^{\lambda}$ coefficients
from Eq. \ref{eq.Legendre}, to using the new \textit{ab initio} PES to carry out RR calculations at $r=r_{eq}$ and our previous
RR calculations which used the CCSD(T) method to compute interaction energies \cite{20MaGiGo}. 
The RR calculations do not carry out the VA procedure of Eq. \ref{eq.vib_coup} and instead expansion coefficients in Eq. \ref{eq.Legendre}
are obtained only for $r = r_{eq}$.

The differences between the VA and RR cross sections obtained using the new PES are small, as is expected since the $\nu=0$ vibrational 
wavefunction is strongly peaked around $r=r_{eq}$. This behaviour was also found for H$_2^+$-He collisions
\cite{17IsGiHe} and justifies our previous treatment of the molecule as a rigid rotor. From Fig. \ref{fig:rot_xsec_comp} it can also be 
seen that the rotationally inelastic collisions using our new PES are in quite good agreement with our previous work as anticipated 
from the similarity of the multipolar expansion coefficients in Figure \ref{fig:V00} to our previous work \cite{20MaGiGo}. The
profiles of the cross sections with energy variation are similar and resonances appear at similar energies. Cross sections only differ 
by tens of percent and these differences will have a negligible effect on the corresponding rotationally inelastic rates.

\begin{figure}[h!]
\centering
\includegraphics[scale=0.33,angle=-90,origin=c]{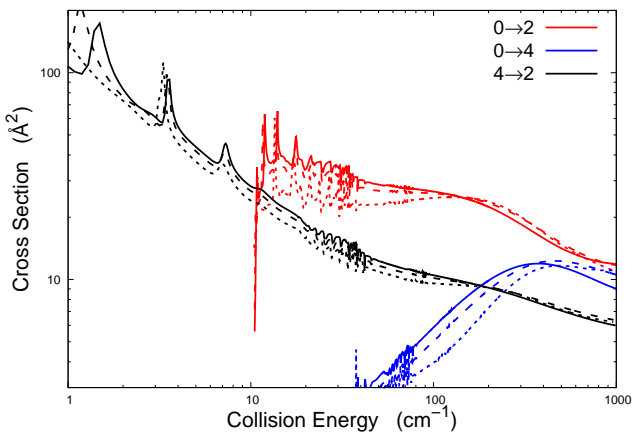}
\vspace*{-15mm}
\caption{Rotationally inelastic vibrationally elastic cross sections for C$_2^-$-He collisions computed for $\nu=0$ (solid lines), 
 $\nu=1$ (long-dashed lines) and $\nu=2$ (short-dashed lines).}
\label{fig:rot_xsec_comp2}
\end{figure}

Rotationally inelastic collisions can also be used to assess the effect of vibrational state on rotationally inelastic collision. 
Fig. \ref{fig:rot_xsec_comp2} shows selected rotationally inelastic vibrationally elastic cross sections 
for C$_2^-$-He for the $\nu = 0, 1$ and 2 vibrational states. As mentioned in Section \ref{sec:3DPES}, 
the $V_{\nu,\nu'}^{\lambda}$ coefficients for $\nu=\nu'$ are very similar resulting in very similar rotationally inelastic cross 
sections for a given vibrational state. This insensitivity of rotationally inelastic cross sections to vibrational state has been seen for
many other molecules undergoing collisions with He \cite{02Krxxxx,06FiSpDh,07FiSpxx,08ToLiKl,17BaDaxx} and Ar  \cite{01KrStxx,01KrMaBu} 
and is a consequence of the small off-diagonal vibrational matrix elements of the PES compared to diagonal ones in Eq. \ref{eq.Legendre}.

The results shown in this section for C$_2^-$-He collisions have demonstrated that rotational and vibrational collisions can essentially
be considered separately. We have recently compared rotationally inelastic thermal quenching cross sections, rates and times for C$_2^-$ in 
collisions with He, Ne and Ar  and we refer the reader to this work for further details of this process \cite{20aMaGiWe}.

The 2D RR PES functions and Legendre expansions for each system are provided in the Supplementary Material \cite{pra_supp}.

\subsection{Vibrationally inelastic cross sections}
\label{sec:vib_xsec}

\begin{figure*}[htb!]
\centering
\includegraphics[scale=0.5,angle=-90,origin=c]{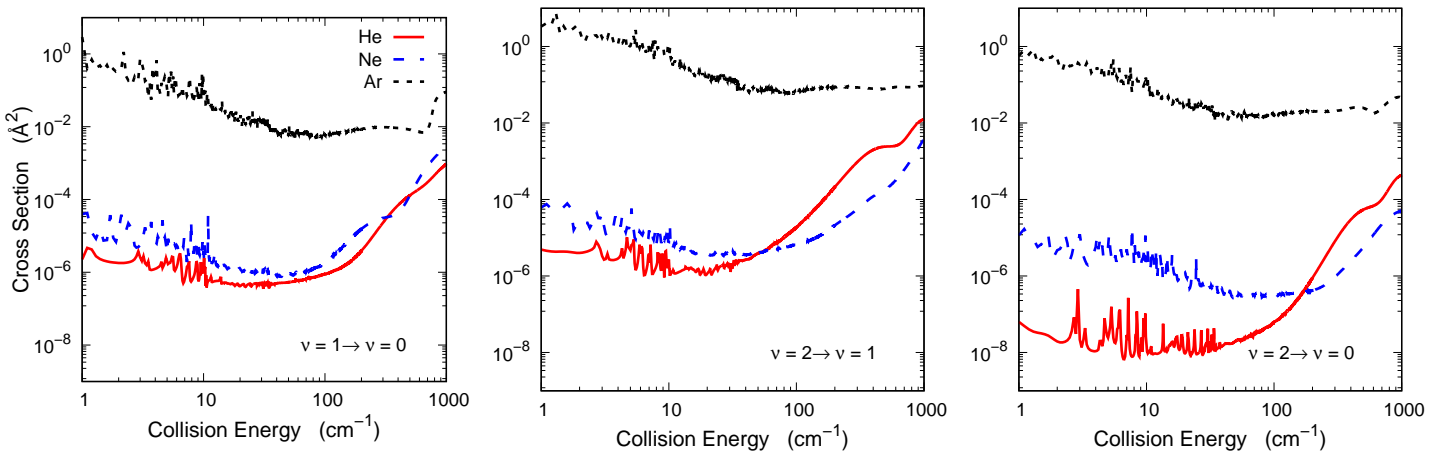}
\vspace*{-70mm}
\caption{Comparison of vibrationally inelastic rotationally elastic cross sections for $\nu=1 \rightarrow \nu=0$ (left panel),
$\nu=2 \rightarrow \nu=1$ (centre panel) and $\nu=2 \rightarrow \nu=0$ (right panel) transitions for collisions of C$_2^-$ with
He, Ne and Ar atoms.}
\label{fig:vib_xsec_comp}
\end{figure*}

As discussed in the Introduction, vibrationally inelastic collisions of C$_2^-$ with helium buffer gas has been suggested to cool
the molecules to their $\nu=0$ vibrational ground state. As shown in Fig. \ref{fig:C2m_PECs}, the ground vibrational state of 
the $A ^2\Pi_u$ electronic state is lower in energy than the $\nu=3$ state of the $X ^2\Sigma_g^+$ ground electronic state.
 It therefore follows that the  vibrational states 
above $\nu = 2$ in $X ^2\Sigma_g^+$ can decay to the $A ^2\Pi_u$ state by dipole allowed transitions which in turn decay into 
the $\nu=2$ and below vibrational states of $X ^2\Sigma_g^+$ \cite{98PeBrAn.c2m}. As a consequence, we only consider here collisional
quenching of the long-lived $\nu=1$ and $\nu=2$ vibrational states of C$_2^-$ in its ground $X ^2\Sigma_g^+$ electronic state. 

From Fig. \ref{fig:C2m_PECs} it can also be seen that the $\nu=2$ vibrational level of the $X ^2\Sigma_g^+$ state is close 
in energy to the $\nu'=0$ state of the excited electronic $A^2\Pi_u$ state with an energy difference relative to the bottom of 
the $X ^2\Sigma_g^+$ PEC of $\approx (4837 - 4380) = 457$ cm$^{-1}$ \cite{16ShLiMe.c2m}. This close energy spacing could cause the
$\nu'=0$ state to perturb the $\nu=2$ state through non-adiabatic effects during collisions. This was also a concern for the neutral
C$_2$-He system where for the isolated molecule the energy difference between the ground $X ^2\Sigma_g^+$ and excited $a^3 \Pi_u$ state
is only around 700 cm$^{-1}$ \cite{19GuJaKr}. For C$_2$-He rigid-rotor rotationally inelastic scattering, Naja \textit{et al.} \cite{08NaAbJa}
found that at the C$_2$ $X ^2\Sigma_g^+$ equilibrium geometry, the energy separation between the $X ^2\Sigma_g^+$ and $a^3 \Pi_u$ states 
is 2000 cm$^{-1}$ which remained at this value even with the approach of the He atom. In this case, electronic state coupling could be 
ignored. When vibrations are involved the situation is more complicated as states can be coupled via the $r$ vibrational coordinate. 
This is the case for the H$^+$-CO and H$^+$-CN systems which have been studied by Kumar \textit{et al.} \cite{10GeKuxx,16AnKuxx,16SaKuxx}.
For these systems the naked positive charge of the proton has a strong perturbing effect and can couple the molecule's electronic states.
The strength of the coupling can be calculated as $\lbrace \psi^{\alpha}_i | \frac{\partial^n }{\partial Q^n } | \psi^{\alpha}_j \rbrace$
where $\psi^{\alpha}_{i/j}$ are the electronic wavefunctions and the operator is the first ($n=1$) or second ($n=2$) derivative 
with respect to the nuclear coordinate $Q$ ($=r$ for diatomics) \cite{10GeKuxx,16AnKuxx,16SaKuxx}. The coupling matrix was 
used to carry out H$^+$-CO scattering calculations
by constructing diabatic PESs for two electronic states allowing a computation of elastic, vibrationally inelastic and charge transfer 
probabilities \cite{11XaKuxx}. The present systems involve the rather weak interaction of an anion with closed-shell noble gases 
so that the gradient couplings between the relevant electronic states should be smaller than in the case of the naked proton as a
partner as in the case of Kumar \textit{et al.'s} work.
Kendrick also recently studied HD + H reactive scattering for excited vibrational states (those energetically 
above and below a conical intersection on the PES) using two coupled diabatic PESs and found significant differences to the 
adiabatic approach \cite{19Kexxxx}. Kendrick has also given a detailed approach for non-adiabatic reactive scattering \cite{18Kexxxx}.
To apply these approaches to the C$_2^-$-He/Ne/Ar systems would involve using an \textit{ab initio} method such as CASSCF or MRCI to 
obtain the ground and electronically excited PESs and assessing them as a function of $R,r$ and $\theta$ coordinates. The coupling
matrices could be calculated as described above and diabatic states constructed. This is somewhat beyond the scope of this work but 
whether non-adiabatic effects would alter quenching rates is an interesting question and would depend on the magnitude of the coupling 
compared to inaccuracies in the adiabatic PES.

Fig. \ref{fig:vib_xsec_comp} compares vibrationally inelastic rotationally elastic (for $j = j' = 0$) cross sections for the
de-excitation $\nu=1 \rightarrow \nu=0$, $\nu=2 \rightarrow \nu=1$ and $\nu=2 \rightarrow \nu=0$ transitions for C$_2^-$ colliding with
He, Ne and Ar atoms. At low collisions energies between 0.1 to around 60 cm$^{-1}$ the cross sections for He and Ne are very small, 
orders of magnitude less than rotationally inelastic cross sections. The $\nu = 2 \rightarrow \nu=0$ process is an order of magnitude 
smaller than the $\Delta \nu = -1$ transitions as is common for cross sections for larger energy differences between states. 
The cross sections show resonances at lower collision energies. These are likely due to shape or Feshbach resonances. 
As is well established, the location and widths of resonances in the scattering cross sections at low collision energies 
are very sensitive to the details of the PES \cite{16KeHaBa,16SuTsxx} and there is currently an efforts to obtain reliable information
about scattering observables, particularly for ultracold regimes, such as the statistical method of Morita \textit{et al.} \cite{19MoKrTs}.
As we are primarily interested in assessing the rates for vibrational quenching for temperatures of 5-100 K, the fine details of the low
energy resonances are less important since the Boltzmann average over the cross sections in this temperature range is likely to be far 
less sensitive to the details of the PES.

For He and Ne, all three processes the cross sections begin to increase in magnitude above 100 cm$^{-1}$. This is a well known trend for
vibrationally inelastic collisions \cite{06FiSpDh,07FiSpxx,08ToLiKl,17BaDaxx}. 
This trend is a consequence of the PES in Fig. \ref{fig:V00} and the vibrational matrix elements in Fig.
\ref{fig:V01}: the off-diagonal matrix elements which couple different vibrational states are only significant at small $R$ values 
where the PES is repulsive. The incoming atom thus requires a higher kinetic energy to allow the scattering wavefunction to become
significant in this region and therefore facilitate vibrational transitions. The C$_2^-$-Ar cross sections are more constant in value at the
higher energies however, a feature which is likely to be due to 
the more attractive potential exhibited by this system, which then allows the scattering wavefunction to build up more significantly in 
the repulsive region, thereby allowing the occurrence of larger inelastic vibrational cross sections.

Comparing the cross sections for collisions of C$_2^-$ with He, Ne and Ar, some general trends are apparent. The cross sections for Ar
are orders of magnitude larger than those of He and Ne which are broadly similar in size. This can be rationalized by comparing the
interactions
shown in Figures \ref{fig:V00} and  \ref{fig:V01}. The well depth of the $V_{0,0}(R,\theta)$ elements for the  C$_2^-$-Ar is far larger than
for He and Ne. This stronger interaction allows the incoming scattering wavefunction to build up more  in the repulsive region where the 
off-diagonal $V_{0,1}(R,\theta)$ are significant, as opposed to He and Ne, for which the coupling potentials are much smaller. 
These  matrix elements are therefore larger for Ar, thus giving rise to larger vibrationally 
inelastic cross sections in comparison with those from the lighter noble gases.

\subsection{Vibrationally inelastic rates}
\label{sec:vib_rates}

\begin{figure*}[htb!]
\centering
\includegraphics[scale=0.5,angle=-90,origin=c]{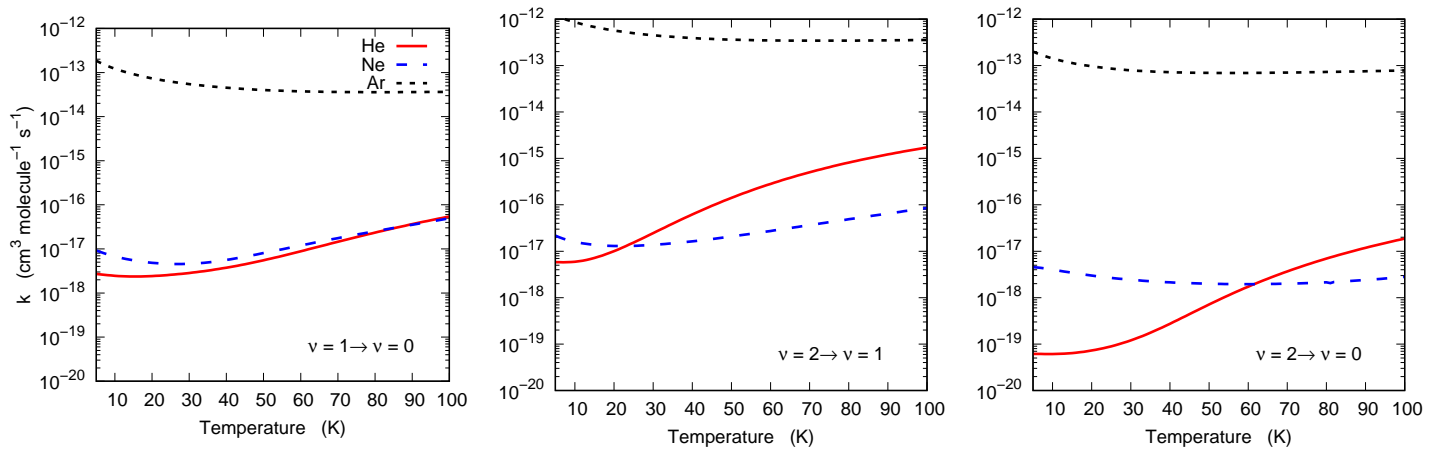}
\vspace*{-70mm}
\caption{Comparison of vibrational rate constants $k_{\nu j \to \nu' j'}(T)$ for $\nu=1 \rightarrow \nu=0$ (left panel),
$\nu=2 \rightarrow \nu=1$ (centre panel) and $\nu=2 \rightarrow \nu=0$ (right panel) transitions for collisions of C$_2^-$ with
He, Ne and Ar atoms}
\label{fig:vib_rates}
\end{figure*}

The computed inelastic cross sections of the previous section can be used to obtain the corresponding thermal rate constants 
$k_{\nu \to \nu'}(T)$, which can be evaluated as the convolution of the computed inelastic cross sections over a Boltzmann 
distribution of the relative collision energies of the interacting partners as
\begin{equation}
k_{\nu \to \nu' }(T) = \left(\displaystyle \frac{8}{\pi \mu k_{B}^3 T^3 } \right)^{1/2} 
 \int_0^{\infty}E_c \sigma_{\nu  \to \nu' }(E_c) e^{-E_c/k_{B}T} dE_c
\label{eq.rateK}
\end{equation}
where $E_c = \mu v^2/2$ is the kinetic energy.

As discussed in the introduction Section, studies on laser cooling of C$_2^-$ have assumed the anion to be  initially cooled to tens of kelvin
\cite{15YzHaGe} and thus the rate constants were computed between 5 and 100 K in 1 K intervals. Fig. \ref{fig:vib_rates} shows 
the rates for vibrationally inelastic rotationally elastic ($j=j'=0$)transitions corresponding to the cross sections in Fig.
\ref{fig:vib_xsec_comp}. The rates for the vibrational de-excitation processes for He and Ne are quite similar, particularly for the $\nu = 1
\to \nu=0$ transition. They increase with increasing temperature as expected from the discussion in Section \ref{sec:vib_xsec} but 
even at 100 K, are at least four orders of magnitude smaller than the rotational de-excitation rates \cite{20aMaGiWe}. For Ar the rates
are more constant and are consistently 2 or 3 orders of magnitude larger than those for He or Ne.
The vibrational quenching rates show some variation with the largest $\nu=2 \rightarrow \nu=1$ rate about an order of magnitude larger 
than the $\nu=1 \rightarrow \nu=0$ and $\nu=2 \rightarrow \nu=0$ rates which themselves are quite similar at the higher 
temperatures considered. 

The trends in rates on going from He to Ne to Ar in collisions with C$_2^-$ are similar to what was found for rotationally inelastic 
collisions \cite{20aMaGiWe} where the stronger interaction with the Ar atom resulted in larger rates and faster themalization times compared
to He or Ne. The larger quenching rates for Ar compared with He and Ne were not easy to predict \textit{a priori}. 
Kato, Bierbaum and Leone measured quenching rates of N$_2^+$ in collisions with He, Ne, Ar, Xe and Kr at 300 K and found quenching rates 
increased with the size of atom \cite{95KaBiLe}. This suggests that the polarizability of the colliding atom plays an 
important role. Ferguson found a similar trend for vibrational quenching of O$_2^+$ with He, Ne and Ar atoms at 300 K \cite{86Fexxxx}. 
In contrast Saidani \textit{et al} calculated  quenching rates for CN with He and Ar over a wide range of temperatures and found that 
cross sections and rates for Ar were orders of magnitude lower than those for He \cite{13SaKaGa}. However, ionic interactions are driven 
by different forces than those acting between neutrals, so it is not obvious how such a result relates to the present findings for an anion.

\begin{table*}[hbt!]
\setlength{\tabcolsep}{10pt}
\renewcommand{\arraystretch}{1.0}
\caption{\label{tab:rates} Comparison of vibrationally inelastic cross sections and rates for different systems. Well depth 
$V_{\mathrm{min}}$ in cm$^{-1}$, cross sections $\sigma_{\nu=1 \to \nu'=0}$ in \AA$^2$ for scattering energy of 10 cm$^{-1}$. 
Temperatures in kelvin and rate constants $k_{\nu=1 \to \nu'=0}(T)$ in cm$^3$ s$^{-1}$. * indicates value was estimated from graph.}
\begin{tabular}{cccccc}
\hline                  
     System   &   $V_{\mathrm{min}}$ & $\sigma_{\nu=1 \to \nu'=0}$ & Temperature & $k_{\nu=1 \to \nu'=0}(T)$ & Reference   \\
\hline
Cations       &                      &                                &             &                           &        \\
 H$_2^+$ + He &  $-$2700             &        10$^*$  &   -           &  -                     & \cite{17IsGiHe} \\
 NO$^+$ + He  &  $-$195              &        1$^*$   &  100          &  7$\times 10^{-14*}$    &  \cite{11StVoxx}       \\
 Mg$^+$ + He  &  $<$ $-$100          &        80$^*$  &   -           &   -                    & \cite{12CaTaGi} \\
 CH$^+$ + He  &   $-$514             &        10$^*$ &   -           &  -                     & \cite{08StVoxx} \\
 BaCl$^+$ + Ca &  $-$7442            &          -    & 0.1           &  1$\times 10^{-9*}$     & \cite{16StHaGa} \\
              &                      &         &                        & \\
Neutrals      &                      &         &                        &        \\
SiO + He      &    $-$27             &   - &   200    &  5$\times 10^{-18*}$   &  \cite{17BaDaxx}   \\
CO + He       &    $-$24             &    1$\times 10^{-8*}$ &  300    &  1$\times 10^{-17}$   &  \cite{02Krxxxx}   \\
SO + He       &    $-$35             &   5$\times 10^{-5*}$         & 300    &  4$\times 10^{-17*}$   &  \cite{06FiSpDh}   \\
CS + He       &    $-$22             &   - &  300    &  1$\times 10^{-17*}$   &  \cite{07FiSpxx}   \\
SiS + He      &    $-$20             &   -& 300    &  4$\times 10^{-17*}$   &  \cite{08ToLiKl}   \\
HF + Ar       &    $-$159            &    1$\times 10^{-6*}$ &       100    &  7$\times 10^{-17}$   &  \cite{01KrMaBu}   \\
CN + He       &    $-$20             &   1$\times 10^{-10*}$  &  100    &  1$\times 10^{-20*}$   &  \cite{13SaKaGa}   \\
CN + Ar       &    $-$130            &   1$\times 10^{-20*}$ & 100  &  1$\times 10^{-28*}$   &  \cite{13SaKaGa}   \\

              &                      &         &                       & \\
Anions        &                      &         &                       & \\  
              &                      &         &                       & \\
C$_2^-$ + He  &   $-30$              & 1.4$\times 10^{-6}$  & 100 & 5.4$\times 10^{-17}$ & This work \\
C$_2^-$ + Ne  &   $-110$             & 3.8$\times 10^{-6}$  & 100 & 4.9$\times 10^{-17}$ & This work \\
C$_2^-$ + Ar  &   $-490$             & 7.2$\times 10^{-2}$  & 100 & 3.7$\times 10^{-14}$ & This work \\
\hline
\end{tabular}
\end{table*}

It is well known that generally, vibrational quenching of molecules due to collisions is inefficient. 
Measurements of low temperature rate constants for quenching of the $\nu = 1$ level include NH undergoing collisions with He 
where the rates are of the order of 4$\times10^{-15}$ cm$^3$ s$^{-1}$ \cite{08CaGrLu}. There are however systems for which collisions 
are efficient at quenching vibrational motion. The low frequency stretching mode (100) in SrOH has been shown to be efficiently de-excited in 
collisions with He \cite{15KoBaMa}. There is also the dramatic case of the BaCl$^+$ + Ca system where laser cooled Ca atoms have been
shown efficiently quench vibrational motion with rates similar to rotational transitions \cite{13ReSuSc,16StHaGa}.

The rates and cross sections calculated here for C$_2^-$-He/Ne/Ar vibrationally inelastic transitions can be compared to other systems.
Table \ref{tab:rates} compares cross sections and rate constants (where available) calculated for various systems for $\nu=1 \rightarrow
\nu=0$ transitions. As a simple comparison metric, Table \ref{tab:rates} also compares the minimum of the interaction potential for each
system, $V_{\mathrm{min}}$. From the table it can be seen that cationic molecules have orders of magnitude larger vibrationally
inelastic cross sections and rates compared with neutral systems. The strength of interaction between the molecule and colliding 
atom, parameterized here by $V_{\mathrm{min}}$, plays a role in the efficiency vibrational quenching as it will allow the scattering
wavefunction to build up for geometries where the coupling matrix is large (Eq. \ref{eq.vib_coup}). This has also been rationalised
in terms of a statistical model \cite{16StHaGa} where the well depth and diatomic vibrational frequency contribute to the density of 
states of the complex which increases the lifetime of the complex and allows efficient quenching. Other factors play a role such
as the mass of the colliding partners (compare the very low rates for CN-Ar collisions compared to CN-He despite the former having
a lower well depth) and the long-range attraction \cite{86Fexxxx}. Nevertheless Table \ref{tab:rates} shows that in general, cations
will have far larger vibrational quenching rates than neutral molecules. The low rates for C$_2^-$-He vibrational quenching are 
similar to those of neutrals as expected based on the relatively weak interaction. To be noted is the behaviour of the 
vibrationally inelastic rates when Ar is taken as the partner gas: the size of the rates brings their values in the range of those 
for cations interacting with He, indicating quantitatively the special behaviour of this noble gas a a partner for C$_2^-$ anions in 
cold traps.

\section{Conclusions}
\label{sec:conc}

The cross sections and corresponding thermal rates for ro-vibrationally inelastic collisions of C$_2^-$ with He, Ne and Ar have been 
calculated using a new  set of \textit{ab initio} PESs. The rotationally inelastic vibrationally elastic cross sections were found to be 
insensitive to vibrational state, justifying the treatment of the C$_2^-$ molecule as a rigid-rotor \cite{20MaGiGo,20aMaGiWe}. 
The cross sections and rates for vibrational quenching from the $\nu=1$ and $\nu=2$ states for He and Ne were found to be orders of 
magnitude lower than those obtained for purely rotationally inelastic collisions. The values of the vibrational quenching rates are found for
this anion to be  similar in size  to those known for other small neutral molecules in collision with helium atoms. For the Ar  partner, 
the vibrationally inelastic rates we have obtained here were around 3 or 4 orders of magnitude larger.

These computed rate coefficients for vibrational quenching can be used to model the behaviour of C$_2^-$ in ion traps with He, Ne or Ar as
buffer gas. It turns out, in fact, that they have significant implications for laser cooling of C$_2^-$: the inefficiency of vibrational
quenching found  for He and Ne in our calculations shows how important the knowledge of vibrational repumping rate coefficients  is for
modelling the cyclic scattering of many photons off C$_2^-$. 
To quench the states which are being populated  within that cycle by using  buffer gas collisions, as  discussed in the present study, 
will require higher pressures in order to efficiently increase the collision frequency in the trap. Our results suggest therefore that argon
would be a more suitable buffer gas to efficiently quench the vibrational motion of C$_2^-$ as lower pressures will be required for it as a
buffer gas in comparison to using either helium or neon.

\section{Acknowledgements}
The C$_2^-$ PEC fit from LEVEL and $\nu = 0, 1$ and 2 vibrational wavefunctions are included in the Supplementary 
Material. Also included are fortran programs for the 2D C$_2^-$-He/Ne/Ar rigid rotor PES and Legendre expansion parameters, 
fortran programs for the 3D C$_2^-$-He/Ne/Ar interaction potential and Legendre expansions of the matrix elements of these surfaces
for $\nu = 0, 1$ and 2 (See equation \ref{eq.vib_elements2}).
We further acknowledge the financial support of the Austrian FWF agency
through research grant n. P29558-N36. One of us (L.G-S) further thanks MINECO (Spain) for grants CTQ2015-65033-P and
PGC2018-09644-B-100. We are grateful for helpful discussions with Graham Worth

\bibliography{c2m_Rg}

\begin{thebibliography}{103}%
\makeatletter
\providecommand \@ifxundefined [1]{%
 \@ifx{#1\undefined}
}%
\providecommand \@ifnum [1]{%
 \ifnum #1\expandafter \@firstoftwo
 \else \expandafter \@secondoftwo
 \fi
}%
\providecommand \@ifx [1]{%
 \ifx #1\expandafter \@firstoftwo
 \else \expandafter \@secondoftwo
 \fi
}%
\providecommand \natexlab [1]{#1}%
\providecommand \enquote  [1]{``#1''}%
\providecommand \bibnamefont  [1]{#1}%
\providecommand \bibfnamefont [1]{#1}%
\providecommand \citenamefont [1]{#1}%
\providecommand \href@noop [0]{\@secondoftwo}%
\providecommand \href [0]{\begingroup \@sanitize@url \@href}%
\providecommand \@href[1]{\@@startlink{#1}\@@href}%
\providecommand \@@href[1]{\endgroup#1\@@endlink}%
\providecommand \@sanitize@url [0]{\catcode `\\12\catcode `\$12\catcode
  `\&12\catcode `\#12\catcode `\^12\catcode `\_12\catcode `\%12\relax}%
\providecommand \@@startlink[1]{}%
\providecommand \@@endlink[0]{}%
\providecommand \url  [0]{\begingroup\@sanitize@url \@url }%
\providecommand \@url [1]{\endgroup\@href {#1}{\urlprefix }}%
\providecommand \urlprefix  [0]{URL }%
\providecommand \Eprint [0]{\href }%
\providecommand \doibase [0]{http://dx.doi.org/}%
\providecommand \selectlanguage [0]{\@gobble}%
\providecommand \bibinfo  [0]{\@secondoftwo}%
\providecommand \bibfield  [0]{\@secondoftwo}%
\providecommand \translation [1]{[#1]}%
\providecommand \BibitemOpen [0]{}%
\providecommand \bibitemStop [0]{}%
\providecommand \bibitemNoStop [0]{.\EOS\space}%
\providecommand \EOS [0]{\spacefactor3000\relax}%
\providecommand \BibitemShut  [1]{\csname bibitem#1\endcsname}%
\let\auto@bib@innerbib\@empty
%</preamble>
\bibitem [{\citenamefont {Tarbutt}(2018)}]{Tarbutt2018:cp}%
  \BibitemOpen
  \bibfield  {author} {\bibinfo {author} {\bibfnamefont {M.}~\bibnamefont
  {Tarbutt}},\ }\href@noop {} {\bibfield  {journal} {\bibinfo  {journal}
  {Contemp. Phys.}\ }\textbf {\bibinfo {volume} {59}},\ \bibinfo {pages} {356}
  (\bibinfo {year} {2018})}\BibitemShut {NoStop}%
\bibitem [{\citenamefont {Barry}\ \emph {et~al.}(2014)\citenamefont {Barry},
  \citenamefont {McCarron}, \citenamefont {Norrgard}, \citenamefont
  {Steinecker},\ and\ \citenamefont {DeMille}}]{Barry2014:n}%
  \BibitemOpen
  \bibfield  {author} {\bibinfo {author} {\bibfnamefont {J.}~\bibnamefont
  {Barry}}, \bibinfo {author} {\bibfnamefont {D.}~\bibnamefont {McCarron}},
  \bibinfo {author} {\bibfnamefont {E.}~\bibnamefont {Norrgard}}, \bibinfo
  {author} {\bibfnamefont {M.}~\bibnamefont {Steinecker}}, \ and\ \bibinfo
  {author} {\bibfnamefont {D.}~\bibnamefont {DeMille}},\ }\href@noop {}
  {\bibfield  {journal} {\bibinfo  {journal} {Nature}\ }\textbf {\bibinfo
  {volume} {512}},\ \bibinfo {pages} {286} (\bibinfo {year}
  {2014})}\BibitemShut {NoStop}%
\bibitem [{\citenamefont {Wolf}\ \emph {et~al.}(2016)\citenamefont {Wolf},
  \citenamefont {Wan}, \citenamefont {Heip}, \citenamefont {Gerbert},
  \citenamefont {Shi},\ and\ \citenamefont {Schmidt}}]{16WoWaHe}%
  \BibitemOpen
  \bibfield  {author} {\bibinfo {author} {\bibfnamefont {F.}~\bibnamefont
  {Wolf}}, \bibinfo {author} {\bibfnamefont {Y.}~\bibnamefont {Wan}}, \bibinfo
  {author} {\bibfnamefont {J.~C.}\ \bibnamefont {Heip}}, \bibinfo {author}
  {\bibfnamefont {F.}~\bibnamefont {Gerbert}}, \bibinfo {author} {\bibfnamefont
  {C.}~\bibnamefont {Shi}}, \ and\ \bibinfo {author} {\bibfnamefont {P.~O.}\
  \bibnamefont {Schmidt}},\ }\href {\doibase doi:10.1038/nature16513}
  {\bibfield  {journal} {\bibinfo  {journal} {Nature}\ }\textbf {\bibinfo
  {volume} {530}},\ \bibinfo {pages} {457} (\bibinfo {year}
  {2016})}\BibitemShut {NoStop}%
\bibitem [{\citenamefont {Loh}\ \emph {et~al.}(2013)\citenamefont {Loh},
  \citenamefont {Cossel}, \citenamefont {Grau}, \citenamefont {Ni},
  \citenamefont {Meyer}, \citenamefont {Bohn}, \citenamefont {Ye},\ and\
  \citenamefont {Cornell}}]{13LoCoGr}%
  \BibitemOpen
  \bibfield  {author} {\bibinfo {author} {\bibfnamefont {H.}~\bibnamefont
  {Loh}}, \bibinfo {author} {\bibfnamefont {K.~C.}\ \bibnamefont {Cossel}},
  \bibinfo {author} {\bibfnamefont {M.~C.}\ \bibnamefont {Grau}}, \bibinfo
  {author} {\bibfnamefont {K.~K.}\ \bibnamefont {Ni}}, \bibinfo {author}
  {\bibfnamefont {E.~R.}\ \bibnamefont {Meyer}}, \bibinfo {author}
  {\bibfnamefont {J.~L.}\ \bibnamefont {Bohn}}, \bibinfo {author}
  {\bibfnamefont {J.}~\bibnamefont {Ye}}, \ and\ \bibinfo {author}
  {\bibfnamefont {E.~A.}\ \bibnamefont {Cornell}},\ }\href {\doibase
  10.1126/science.1243683} {\bibfield  {journal} {\bibinfo  {journal}
  {Science}\ }\textbf {\bibinfo {volume} {342}},\ \bibinfo {pages} {1220}
  (\bibinfo {year} {2013})}\BibitemShut {NoStop}%
\bibitem [{\citenamefont {D\"{o}rfler}\ \emph {et~al.}(2019)\citenamefont
  {D\"{o}rfler}, \citenamefont {Eberle}, \citenamefont {Koner}, \citenamefont
  {Tomza}, \citenamefont {Meuwly},\ and\ \citenamefont {Willitsch}}]{19DoEbKo}%
  \BibitemOpen
  \bibfield  {author} {\bibinfo {author} {\bibfnamefont {A.~D.}\ \bibnamefont
  {D\"{o}rfler}}, \bibinfo {author} {\bibfnamefont {P.}~\bibnamefont {Eberle}},
  \bibinfo {author} {\bibfnamefont {D.}~\bibnamefont {Koner}}, \bibinfo
  {author} {\bibfnamefont {M.}~\bibnamefont {Tomza}}, \bibinfo {author}
  {\bibfnamefont {M.}~\bibnamefont {Meuwly}}, \ and\ \bibinfo {author}
  {\bibfnamefont {S.}~\bibnamefont {Willitsch}},\ }\href {\doibase
  10.1038/s41467-019-13218-x} {\bibfield  {journal} {\bibinfo  {journal} {Nat.
  Commun.}\ }\textbf {\bibinfo {volume} {10}},\ \bibinfo {pages} {5429}
  (\bibinfo {year} {2019})}\BibitemShut {NoStop}%
\bibitem [{\citenamefont {Yzombard}\ \emph {et~al.}(2015)\citenamefont
  {Yzombard}, \citenamefont {Hamamda}, \citenamefont {Gerber}, \citenamefont
  {Doser},\ and\ \citenamefont {Comparat}}]{15YzHaGe}%
  \BibitemOpen
  \bibfield  {author} {\bibinfo {author} {\bibfnamefont {P.}~\bibnamefont
  {Yzombard}}, \bibinfo {author} {\bibfnamefont {M.}~\bibnamefont {Hamamda}},
  \bibinfo {author} {\bibfnamefont {S.}~\bibnamefont {Gerber}}, \bibinfo
  {author} {\bibfnamefont {M.}~\bibnamefont {Doser}}, \ and\ \bibinfo {author}
  {\bibfnamefont {D.}~\bibnamefont {Comparat}},\ }\href {\doibase
  10.1103/PhysRevLett.114.213001} {\bibfield  {journal} {\bibinfo  {journal}
  {Phys. Rev. Lett.}\ }\textbf {\bibinfo {volume} {114}},\ \bibinfo {pages}
  {213001} (\bibinfo {year} {2015})}\BibitemShut {NoStop}%
\bibitem [{\citenamefont {Shan-Shan}\ \emph {et~al.}(2003)\citenamefont
  {Shan-Shan}, \citenamefont {Xiao-Hua}, \citenamefont {Ben-Xi}, \citenamefont
  {Kakule}, \citenamefont {Sheng-Hai}, \citenamefont {Ying-Chun}, \citenamefont
  {Yu-Yan},\ and\ \citenamefont {Yang-Qin}}]{03ShXiBe}%
  \BibitemOpen
  \bibfield  {author} {\bibinfo {author} {\bibfnamefont {Y.}~\bibnamefont
  {Shan-Shan}}, \bibinfo {author} {\bibfnamefont {Y.}~\bibnamefont {Xiao-Hua}},
  \bibinfo {author} {\bibfnamefont {L.}~\bibnamefont {Ben-Xi}}, \bibinfo
  {author} {\bibfnamefont {K.}~\bibnamefont {Kakule}}, \bibinfo {author}
  {\bibfnamefont {W.}~\bibnamefont {Sheng-Hai}}, \bibinfo {author}
  {\bibfnamefont {G.}~\bibnamefont {Ying-Chun}}, \bibinfo {author}
  {\bibfnamefont {L.}~\bibnamefont {Yu-Yan}}, \ and\ \bibinfo {author}
  {\bibfnamefont {C.}~\bibnamefont {Yang-Qin}},\ }\href {\doibase
  10.1088/1009-1963/12/7/308} {\bibfield  {journal} {\bibinfo  {journal}
  {Chinese Phys.}\ }\textbf {\bibinfo {volume} {12}},\ \bibinfo {pages} {745}
  (\bibinfo {year} {2003})}\BibitemShut {NoStop}%
\bibitem [{\citenamefont {Shi}\ \emph {et~al.}(2016)\citenamefont {Shi},
  \citenamefont {Li}, \citenamefont {Meng}, \citenamefont {Wei}, \citenamefont
  {Deng},\ and\ \citenamefont {Yang}}]{16ShLiMe.c2m}%
  \BibitemOpen
  \bibfield  {author} {\bibinfo {author} {\bibfnamefont {W.}~\bibnamefont
  {Shi}}, \bibinfo {author} {\bibfnamefont {C.}~\bibnamefont {Li}}, \bibinfo
  {author} {\bibfnamefont {H.}~\bibnamefont {Meng}}, \bibinfo {author}
  {\bibfnamefont {J.}~\bibnamefont {Wei}}, \bibinfo {author} {\bibfnamefont
  {L.}~\bibnamefont {Deng}}, \ and\ \bibinfo {author} {\bibfnamefont
  {C.}~\bibnamefont {Yang}},\ }\href {\doibase 10.1016/j.comptc.2016.01.015}
  {\bibfield  {journal} {\bibinfo  {journal} {Comput. Theor. Chem.}\ }\textbf
  {\bibinfo {volume} {1079}},\ \bibinfo {pages} {57} (\bibinfo {year}
  {2016})}\BibitemShut {NoStop}%
\bibitem [{\citenamefont {Fesel}\ \emph {et~al.}(2017)\citenamefont {Fesel},
  \citenamefont {Gerber}, \citenamefont {Doser},\ and\ \citenamefont
  {Comparat}}]{17FeGeDo}%
  \BibitemOpen
  \bibfield  {author} {\bibinfo {author} {\bibfnamefont {J.}~\bibnamefont
  {Fesel}}, \bibinfo {author} {\bibfnamefont {S.}~\bibnamefont {Gerber}},
  \bibinfo {author} {\bibfnamefont {M.}~\bibnamefont {Doser}}, \ and\ \bibinfo
  {author} {\bibfnamefont {D.}~\bibnamefont {Comparat}},\ }\href {\doibase
  10.1103/PhysRevA.96.031401} {\bibfield  {journal} {\bibinfo  {journal} {Phys.
  Rev. A}\ }\textbf {\bibinfo {volume} {96}},\ \bibinfo {pages} {031401(R)}
  (\bibinfo {year} {2017})}\BibitemShut {NoStop}%
\bibitem [{\citenamefont {Gerber}\ \emph {et~al.}(2018)\citenamefont {Gerber},
  \citenamefont {Fesel}, \citenamefont {Doser},\ and\ \citenamefont
  {Comparat}}]{18GeFeDo}%
  \BibitemOpen
  \bibfield  {author} {\bibinfo {author} {\bibfnamefont {S.}~\bibnamefont
  {Gerber}}, \bibinfo {author} {\bibfnamefont {J.}~\bibnamefont {Fesel}},
  \bibinfo {author} {\bibfnamefont {M.}~\bibnamefont {Doser}}, \ and\ \bibinfo
  {author} {\bibfnamefont {D.}~\bibnamefont {Comparat}},\ }\href {\doibase
  10.1088/1367-2630/aaa951} {\bibfield  {journal} {\bibinfo  {journal} {New J.
  Phys.}\ }\textbf {\bibinfo {volume} {20}},\ \bibinfo {pages} {023024}
  (\bibinfo {year} {2018})}\BibitemShut {NoStop}%
\bibitem [{\citenamefont {Ahmadi~et al.}(2017)}]{17Ahxxxx}%
  \BibitemOpen
  \bibfield  {author} {\bibinfo {author} {\bibfnamefont {M.}~\bibnamefont
  {Ahmadi~et al.}},\ }\href {\doibase 10.1038/nature21040} {\bibfield
  {journal} {\bibinfo  {journal} {Nature}\ }\textbf {\bibinfo {volume} {541}},\
  \bibinfo {pages} {506} (\bibinfo {year} {2017})}\BibitemShut {NoStop}%
\bibitem [{\citenamefont {Perez}\ and\ \citenamefont
  {Sacquin}(2012)}]{12PeSaxx}%
  \BibitemOpen
  \bibfield  {author} {\bibinfo {author} {\bibfnamefont {P.}~\bibnamefont
  {Perez}}\ and\ \bibinfo {author} {\bibfnamefont {Y.}~\bibnamefont
  {Sacquin}},\ }\href {\doibase 10.1088/0264-9381/29/18/184008} {\bibfield
  {journal} {\bibinfo  {journal} {Class. Quantum. Grav.}\ }\textbf {\bibinfo
  {volume} {29}},\ \bibinfo {pages} {184008} (\bibinfo {year}
  {2012})}\BibitemShut {NoStop}%
\bibitem [{\citenamefont {Herzberg}\ and\ \citenamefont
  {Lagerqvist}(1968)}]{68HeLaxx.c2m}%
  \BibitemOpen
  \bibfield  {author} {\bibinfo {author} {\bibfnamefont {G.}~\bibnamefont
  {Herzberg}}\ and\ \bibinfo {author} {\bibfnamefont {A.}~\bibnamefont
  {Lagerqvist}},\ }\href {\doibase 10.1139/p68-596} {\bibfield  {journal}
  {\bibinfo  {journal} {Can. J. Phys.}\ }\textbf {\bibinfo {volume} {46}},\
  \bibinfo {pages} {2363} (\bibinfo {year} {1968})}\BibitemShut {NoStop}%
\bibitem [{\citenamefont {Milligan}\ and\ \citenamefont
  {Jacox}(1969)}]{69MiMaxx.c2m}%
  \BibitemOpen
  \bibfield  {author} {\bibinfo {author} {\bibfnamefont {E.~D.}\ \bibnamefont
  {Milligan}}\ and\ \bibinfo {author} {\bibfnamefont {M.~E.}\ \bibnamefont
  {Jacox}},\ }\href {\doibase 10.1063/1.1672283} {\bibfield  {journal}
  {\bibinfo  {journal} {J. Chem. Phys}\ }\textbf {\bibinfo {volume} {51}},\
  \bibinfo {pages} {1952} (\bibinfo {year} {1969})}\BibitemShut {NoStop}%
\bibitem [{\citenamefont {Frosch}(1971)}]{71.Frxxxx.c2m}%
  \BibitemOpen
  \bibfield  {author} {\bibinfo {author} {\bibfnamefont {R.~P.}\ \bibnamefont
  {Frosch}},\ }\href {\doibase 10.1063/1.1675229} {\bibfield  {journal}
  {\bibinfo  {journal} {J. Chem. Phys.}\ }\textbf {\bibinfo {volume} {54}},\
  \bibinfo {pages} {2660} (\bibinfo {year} {1971})}\BibitemShut {NoStop}%
\bibitem [{\citenamefont {Lineberger}\ and\ \citenamefont
  {Patterson}(1972)}]{72LiPaxx.c2m}%
  \BibitemOpen
  \bibfield  {author} {\bibinfo {author} {\bibfnamefont {W.~C.}\ \bibnamefont
  {Lineberger}}\ and\ \bibinfo {author} {\bibfnamefont {T.~A.}\ \bibnamefont
  {Patterson}},\ }\href {\doibase 10.1016/0009-2614(72)80037-X} {\bibfield
  {journal} {\bibinfo  {journal} {Chem. Phys. Lett.}\ }\textbf {\bibinfo
  {volume} {13}},\ \bibinfo {pages} {40} (\bibinfo {year} {1972})}\BibitemShut
  {NoStop}%
\bibitem [{\citenamefont {Jones}\ \emph {et~al.}(1980)\citenamefont {Jones},
  \citenamefont {Mead}, \citenamefont {Kohler}, \citenamefont {Rosner},\ and\
  \citenamefont {Lineberger}}]{80JoMeKo.c2m}%
  \BibitemOpen
  \bibfield  {author} {\bibinfo {author} {\bibfnamefont {P.~L.}\ \bibnamefont
  {Jones}}, \bibinfo {author} {\bibfnamefont {R.~D.}\ \bibnamefont {Mead}},
  \bibinfo {author} {\bibfnamefont {B.~E.}\ \bibnamefont {Kohler}}, \bibinfo
  {author} {\bibfnamefont {S.~D.}\ \bibnamefont {Rosner}}, \ and\ \bibinfo
  {author} {\bibfnamefont {W.~C.}\ \bibnamefont {Lineberger}},\ }\href
  {\doibase 10.1063/1.440678} {\bibfield  {journal} {\bibinfo  {journal} {J.
  Chem. Phys.}\ }\textbf {\bibinfo {volume} {73}},\ \bibinfo {pages} {4419}
  (\bibinfo {year} {1980})}\BibitemShut {NoStop}%
\bibitem [{\citenamefont {Leutwyler}\ \emph {et~al.}(1982)\citenamefont
  {Leutwyler}, \citenamefont {Maier},\ and\ \citenamefont
  {Misev}}]{82LeNaMi.c2m}%
  \BibitemOpen
  \bibfield  {author} {\bibinfo {author} {\bibfnamefont {S.}~\bibnamefont
  {Leutwyler}}, \bibinfo {author} {\bibfnamefont {J.~P.}\ \bibnamefont
  {Maier}}, \ and\ \bibinfo {author} {\bibfnamefont {L.}~\bibnamefont
  {Misev}},\ }\href {\doibase 10.1016/0009-2614(82)83642-7} {\bibfield
  {journal} {\bibinfo  {journal} {Chem. Phys. Lett.}\ }\textbf {\bibinfo
  {volume} {91}},\ \bibinfo {pages} {206} (\bibinfo {year} {1982})}\BibitemShut
  {NoStop}%
\bibitem [{\citenamefont {Mead}\ \emph {et~al.}(1985)\citenamefont {Mead},
  \citenamefont {Hefter}, \citenamefont {Schulz},\ and\ \citenamefont
  {Lineberger}}]{85MeHeSc.c2m}%
  \BibitemOpen
  \bibfield  {author} {\bibinfo {author} {\bibfnamefont {R.~D.}\ \bibnamefont
  {Mead}}, \bibinfo {author} {\bibfnamefont {U.}~\bibnamefont {Hefter}},
  \bibinfo {author} {\bibfnamefont {P.~A.}\ \bibnamefont {Schulz}}, \ and\
  \bibinfo {author} {\bibfnamefont {W.~C.}\ \bibnamefont {Lineberger}},\ }\href
  {\doibase 10.1063/1.448960} {\bibfield  {journal} {\bibinfo  {journal} {J.
  Chem. Phys.}\ }\textbf {\bibinfo {volume} {82}},\ \bibinfo {pages} {1723}
  (\bibinfo {year} {1985})}\BibitemShut {NoStop}%
\bibitem [{\citenamefont {Rehfuss}\ \emph {et~al.}(1988)\citenamefont
  {Rehfuss}, \citenamefont {Liu}, \citenamefont {Dinelli}, \citenamefont
  {Jagod}, \citenamefont {Ho}, \citenamefont {Crofton},\ and\ \citenamefont
  {Oka}}]{88ReLiDi.c2m}%
  \BibitemOpen
  \bibfield  {author} {\bibinfo {author} {\bibfnamefont {B.~D.}\ \bibnamefont
  {Rehfuss}}, \bibinfo {author} {\bibfnamefont {D.-J.}\ \bibnamefont {Liu}},
  \bibinfo {author} {\bibfnamefont {B.~M.}\ \bibnamefont {Dinelli}}, \bibinfo
  {author} {\bibfnamefont {M.-F.}\ \bibnamefont {Jagod}}, \bibinfo {author}
  {\bibfnamefont {W.~C.}\ \bibnamefont {Ho}}, \bibinfo {author} {\bibfnamefont
  {M.~W.}\ \bibnamefont {Crofton}}, \ and\ \bibinfo {author} {\bibfnamefont
  {T.}~\bibnamefont {Oka}},\ }\href@noop {} {\bibfield  {journal} {\bibinfo
  {journal} {J. Chem. Phys}\ }\textbf {\bibinfo {volume} {89}},\ \bibinfo
  {pages} {129} (\bibinfo {year} {1988})}\BibitemShut {NoStop}%
\bibitem [{\citenamefont {Ervin}\ and\ \citenamefont
  {Lineberger}(1991)}]{91ErLixx.c2m}%
  \BibitemOpen
  \bibfield  {author} {\bibinfo {author} {\bibfnamefont {K.~M.}\ \bibnamefont
  {Ervin}}\ and\ \bibinfo {author} {\bibfnamefont {W.~C.}\ \bibnamefont
  {Lineberger}},\ }\href {\doibase 10.1021/j100156a026} {\bibfield  {journal}
  {\bibinfo  {journal} {J. Phys. Chem.}\ }\textbf {\bibinfo {volume} {95}},\
  \bibinfo {pages} {1167} (\bibinfo {year} {1991})}\BibitemShut {NoStop}%
\bibitem [{\citenamefont {Royen}\ and\ \citenamefont
  {Zackrisson}(1992)}]{92RoZaxx.c2m}%
  \BibitemOpen
  \bibfield  {author} {\bibinfo {author} {\bibfnamefont {P.}~\bibnamefont
  {Royen}}\ and\ \bibinfo {author} {\bibfnamefont {M.}~\bibnamefont
  {Zackrisson}},\ }\href {\doibase 10.1016/0022-2852(92)90534-U} {\bibfield
  {journal} {\bibinfo  {journal} {J. Mol. Spectrosc.}\ }\textbf {\bibinfo
  {volume} {155}},\ \bibinfo {pages} {427} (\bibinfo {year}
  {1992})}\BibitemShut {NoStop}%
\bibitem [{\citenamefont {Beer}\ \emph {et~al.}(1995)\citenamefont {Beer},
  \citenamefont {Zhao}, \citenamefont {Yourshaw},\ and\ \citenamefont
  {Neumark}}]{95BeZhYu.c2m}%
  \BibitemOpen
  \bibfield  {author} {\bibinfo {author} {\bibfnamefont {E.~d.}\ \bibnamefont
  {Beer}}, \bibinfo {author} {\bibfnamefont {Y.}~\bibnamefont {Zhao}}, \bibinfo
  {author} {\bibfnamefont {I.}~\bibnamefont {Yourshaw}}, \ and\ \bibinfo
  {author} {\bibfnamefont {D.~M.}\ \bibnamefont {Neumark}},\ }\href {\doibase
  10.1016/0009-2614(95)00967-9} {\bibfield  {journal} {\bibinfo  {journal}
  {Chem. Phys. Lett.}\ }\textbf {\bibinfo {volume} {244}},\ \bibinfo {pages}
  {400} (\bibinfo {year} {1995})}\BibitemShut {NoStop}%
\bibitem [{\citenamefont {Pedersen}\ \emph {et~al.}(1998)\citenamefont
  {Pedersen}, \citenamefont {Brink}, \citenamefont {H}, \citenamefont {Bjerre},
  \citenamefont {Hvelplund}, \citenamefont {Kella},\ and\ \citenamefont
  {Shen}}]{98PeBrAn.c2m}%
  \BibitemOpen
  \bibfield  {author} {\bibinfo {author} {\bibfnamefont {H.~B.}\ \bibnamefont
  {Pedersen}}, \bibinfo {author} {\bibfnamefont {C.}~\bibnamefont {Brink}},
  \bibinfo {author} {\bibfnamefont {A.~L.}\ \bibnamefont {H}}, \bibinfo
  {author} {\bibfnamefont {N.}~\bibnamefont {Bjerre}}, \bibinfo {author}
  {\bibfnamefont {P.}~\bibnamefont {Hvelplund}}, \bibinfo {author}
  {\bibfnamefont {D.}~\bibnamefont {Kella}}, \ and\ \bibinfo {author}
  {\bibfnamefont {H.}~\bibnamefont {Shen}},\ }\href {\doibase 10.1063/1.477207}
  {\bibfield  {journal} {\bibinfo  {journal} {J. Chem. Phys.}\ }\textbf
  {\bibinfo {volume} {464}},\ \bibinfo {pages} {5849} (\bibinfo {year}
  {1998})}\BibitemShut {NoStop}%
\bibitem [{\citenamefont {Bragg}\ \emph {et~al.}(2003)\citenamefont {Bragg},
  \citenamefont {Wester}, \citenamefont {Davis}, \citenamefont {Kammrath},\
  and\ \citenamefont {Neumark}}]{03BrWeDa.c2m}%
  \BibitemOpen
  \bibfield  {author} {\bibinfo {author} {\bibfnamefont {A.~E.}\ \bibnamefont
  {Bragg}}, \bibinfo {author} {\bibfnamefont {R.}~\bibnamefont {Wester}},
  \bibinfo {author} {\bibfnamefont {A.~V.}\ \bibnamefont {Davis}}, \bibinfo
  {author} {\bibfnamefont {A.}~\bibnamefont {Kammrath}}, \ and\ \bibinfo
  {author} {\bibfnamefont {D.~M.}\ \bibnamefont {Neumark}},\ }\href {\doibase
  10.1016/S0009-2614(03)01060-1} {\bibfield  {journal} {\bibinfo  {journal}
  {Chem. Phys. Lett.}\ }\textbf {\bibinfo {volume} {376}},\ \bibinfo {pages}
  {767} (\bibinfo {year} {2003})}\BibitemShut {NoStop}%
\bibitem [{\citenamefont {Nakajima}(2017)}]{17Naxxxx.c2m}%
  \BibitemOpen
  \bibfield  {author} {\bibinfo {author} {\bibfnamefont {M.}~\bibnamefont
  {Nakajima}},\ }\href {\doibase 10.1016/j.jms.2016.11.002} {\bibfield
  {journal} {\bibinfo  {journal} {J. Mol. Spectrosc.}\ }\textbf {\bibinfo
  {volume} {331}},\ \bibinfo {pages} {106} (\bibinfo {year}
  {2017})}\BibitemShut {NoStop}%
\bibitem [{\citenamefont {Endres}\ \emph {et~al.}(2014)\citenamefont {Endres},
  \citenamefont {Lakhmanskaya}, \citenamefont {Hauser}, \citenamefont {Huber},
  \citenamefont {Best}, \citenamefont {Kumar}, \citenamefont {Probst},\ and\
  \citenamefont {Wester}}]{14EnLaHa.c2m}%
  \BibitemOpen
  \bibfield  {author} {\bibinfo {author} {\bibfnamefont {E.~S.}\ \bibnamefont
  {Endres}}, \bibinfo {author} {\bibfnamefont {O.}~\bibnamefont
  {Lakhmanskaya}}, \bibinfo {author} {\bibfnamefont {D.}~\bibnamefont
  {Hauser}}, \bibinfo {author} {\bibfnamefont {S.~E.}\ \bibnamefont {Huber}},
  \bibinfo {author} {\bibfnamefont {T.}~\bibnamefont {Best}}, \bibinfo {author}
  {\bibfnamefont {S.~S.}\ \bibnamefont {Kumar}}, \bibinfo {author}
  {\bibfnamefont {M.}~\bibnamefont {Probst}}, \ and\ \bibinfo {author}
  {\bibfnamefont {R.}~\bibnamefont {Wester}},\ }\href {\doibase
  10.1021/jp504242p} {\bibfield  {journal} {\bibinfo  {journal} {J. Phys. Chem.
  A}\ }\textbf {\bibinfo {volume} {118}},\ \bibinfo {pages} {6705} (\bibinfo
  {year} {2014})}\BibitemShut {NoStop}%
\bibitem [{\citenamefont {Barsuhn}(1974)}]{74Baxxxx.c2m}%
  \BibitemOpen
  \bibfield  {author} {\bibinfo {author} {\bibfnamefont {J.}~\bibnamefont
  {Barsuhn}},\ }\href {\doibase 10.1088/0022-3700/7/1/025} {\bibfield
  {journal} {\bibinfo  {journal} {J. Phys. B: At., Mol. Opt. Phys}\ }\textbf
  {\bibinfo {volume} {7}},\ \bibinfo {pages} {155} (\bibinfo {year}
  {1974})}\BibitemShut {NoStop}%
\bibitem [{\citenamefont {Zeitz}\ \emph {et~al.}(1979)\citenamefont {Zeitz},
  \citenamefont {Peyerimhoff},\ and\ \citenamefont {Buenker}}]{79ZePeBu.c2m}%
  \BibitemOpen
  \bibfield  {author} {\bibinfo {author} {\bibfnamefont {M.}~\bibnamefont
  {Zeitz}}, \bibinfo {author} {\bibfnamefont {S.~D.}\ \bibnamefont
  {Peyerimhoff}}, \ and\ \bibinfo {author} {\bibfnamefont {R.~J.}\ \bibnamefont
  {Buenker}},\ }\href {\doibase 10.1016/0009-2614(79)80505-9} {\bibfield
  {journal} {\bibinfo  {journal} {Chem. Phys. Lett.}\ }\textbf {\bibinfo
  {volume} {64}},\ \bibinfo {pages} {243} (\bibinfo {year} {1979})}\BibitemShut
  {NoStop}%
\bibitem [{\citenamefont {Dupuis}\ and\ \citenamefont
  {Liu}(1980)}]{80DuLixx.c2m}%
  \BibitemOpen
  \bibfield  {author} {\bibinfo {author} {\bibfnamefont {M.}~\bibnamefont
  {Dupuis}}\ and\ \bibinfo {author} {\bibfnamefont {B.}~\bibnamefont {Liu}},\
  }\href {\doibase 10.1063/1.439879} {\bibfield  {journal} {\bibinfo  {journal}
  {J. Chem. Phys.}\ }\textbf {\bibinfo {volume} {73}},\ \bibinfo {pages} {337}
  (\bibinfo {year} {1980})}\BibitemShut {NoStop}%
\bibitem [{\citenamefont {Rosmus}\ and\ \citenamefont
  {Werner}(1984)}]{84RoWexx.c2m}%
  \BibitemOpen
  \bibfield  {author} {\bibinfo {author} {\bibfnamefont {P.}~\bibnamefont
  {Rosmus}}\ and\ \bibinfo {author} {\bibfnamefont {H.-J.}\ \bibnamefont
  {Werner}},\ }\href {\doibase 10.1063/1.446579} {\bibfield  {journal}
  {\bibinfo  {journal} {J. Chem. Phys.}\ }\textbf {\bibinfo {volume} {80}},\
  \bibinfo {pages} {5085} (\bibinfo {year} {1984})}\BibitemShut {NoStop}%
\bibitem [{\citenamefont {Nichols}\ and\ \citenamefont
  {Simons}(1987)}]{87NiSixx.c2m}%
  \BibitemOpen
  \bibfield  {author} {\bibinfo {author} {\bibfnamefont {J.~A.}\ \bibnamefont
  {Nichols}}\ and\ \bibinfo {author} {\bibfnamefont {J.}~\bibnamefont
  {Simons}},\ }\href {\doibase 10.1063/1.452345} {\bibfield  {journal}
  {\bibinfo  {journal} {J. Chem. Phys.}\ }\textbf {\bibinfo {volume} {86}},\
  \bibinfo {pages} {6972} (\bibinfo {year} {1987})}\BibitemShut {NoStop}%
\bibitem [{\citenamefont {Watts}\ and\ \citenamefont
  {Bertlett}(1992)}]{92WaBaxx.c2m}%
  \BibitemOpen
  \bibfield  {author} {\bibinfo {author} {\bibfnamefont {J.~D.}\ \bibnamefont
  {Watts}}\ and\ \bibinfo {author} {\bibfnamefont {R.~J.}\ \bibnamefont
  {Bertlett}},\ }\href {\doibase 10.1063/1.4662649} {\bibfield  {journal}
  {\bibinfo  {journal} {J. Chem. Phys.}\ }\textbf {\bibinfo {volume} {96}},\
  \bibinfo {pages} {6073} (\bibinfo {year} {1992})}\BibitemShut {NoStop}%
\bibitem [{\citenamefont {\u{S}edivcov\'{a}}\ and\ \citenamefont
  {\u{S}pirko}(2006)}]{06SeSpxx.c2m}%
  \BibitemOpen
  \bibfield  {author} {\bibinfo {author} {\bibfnamefont {T.}~\bibnamefont
  {\u{S}edivcov\'{a}}}\ and\ \bibinfo {author} {\bibfnamefont {V.}~\bibnamefont
  {\u{S}pirko}},\ }\href {\doibase 10.1080/00268970600662689} {\bibfield
  {journal} {\bibinfo  {journal} {Mol. Phys.}\ }\textbf {\bibinfo {volume}
  {104}},\ \bibinfo {pages} {1999} (\bibinfo {year} {2006})}\BibitemShut
  {NoStop}%
\bibitem [{\citenamefont {Kas}\ \emph {et~al.}(2019)\citenamefont {Kas},
  \citenamefont {Loreau}, \citenamefont {Li\'{e}vin},\ and\ \citenamefont
  {Vaeck}}]{19KaLoLi.c2m}%
  \BibitemOpen
  \bibfield  {author} {\bibinfo {author} {\bibfnamefont {M.}~\bibnamefont
  {Kas}}, \bibinfo {author} {\bibfnamefont {J.}~\bibnamefont {Loreau}},
  \bibinfo {author} {\bibfnamefont {J.}~\bibnamefont {Li\'{e}vin}}, \ and\
  \bibinfo {author} {\bibfnamefont {N.}~\bibnamefont {Vaeck}},\ }\href
  {\doibase 10.1103/PhysRevA.99.042702} {\bibfield  {journal} {\bibinfo
  {journal} {Phys. Rev. A}\ }\textbf {\bibinfo {volume} {99}},\ \bibinfo
  {pages} {042702} (\bibinfo {year} {2019})}\BibitemShut {NoStop}%
\bibitem [{\citenamefont {Gulania}\ \emph {et~al.}(2019)\citenamefont
  {Gulania}, \citenamefont {Jagau},\ and\ \citenamefont {Krylov}}]{19GuJaKr}%
  \BibitemOpen
  \bibfield  {author} {\bibinfo {author} {\bibfnamefont {S.}~\bibnamefont
  {Gulania}}, \bibinfo {author} {\bibfnamefont {T.-C.}\ \bibnamefont {Jagau}},
  \ and\ \bibinfo {author} {\bibfnamefont {A.~I.}\ \bibnamefont {Krylov}},\
  }\href {\doibase 10.1039/c8fd00185e} {\bibfield  {journal} {\bibinfo
  {journal} {Farady Discuss.}\ }\textbf {\bibinfo {volume} {217}},\ \bibinfo
  {pages} {514} (\bibinfo {year} {2019})}\BibitemShut {NoStop}%
\bibitem [{\citenamefont {Lambert}\ \emph {et~al.}(1995)\citenamefont
  {Lambert}, \citenamefont {Sheffer},\ and\ \citenamefont
  {Federman}}]{95LaShFe.c2m}%
  \BibitemOpen
  \bibfield  {author} {\bibinfo {author} {\bibfnamefont {D.~L.}\ \bibnamefont
  {Lambert}}, \bibinfo {author} {\bibfnamefont {Y.}~\bibnamefont {Sheffer}}, \
  and\ \bibinfo {author} {\bibfnamefont {S.~R.}\ \bibnamefont {Federman}},\
  }\href {\doibase 10.1086/175119} {\bibfield  {journal} {\bibinfo  {journal}
  {Astrophys J.}\ }\textbf {\bibinfo {volume} {438}},\ \bibinfo {pages} {740}
  (\bibinfo {year} {1995})}\BibitemShut {NoStop}%
\bibitem [{\citenamefont {Lambert}\ \emph {et~al.}(1990)\citenamefont
  {Lambert}, \citenamefont {Sheffer}, \citenamefont {Danks}, \citenamefont
  {Arpigny},\ and\ \citenamefont {Magain}}]{90LaShDa.c2m}%
  \BibitemOpen
  \bibfield  {author} {\bibinfo {author} {\bibfnamefont {D.~L.}\ \bibnamefont
  {Lambert}}, \bibinfo {author} {\bibfnamefont {Y.}~\bibnamefont {Sheffer}},
  \bibinfo {author} {\bibfnamefont {A.~C.}\ \bibnamefont {Danks}}, \bibinfo
  {author} {\bibfnamefont {C.}~\bibnamefont {Arpigny}}, \ and\ \bibinfo
  {author} {\bibfnamefont {P.}~\bibnamefont {Magain}},\ }\href {\doibase
  10.1086/168654} {\bibfield  {journal} {\bibinfo  {journal} {Astrophys J.}\
  }\textbf {\bibinfo {volume} {353}},\ \bibinfo {pages} {640} (\bibinfo {year}
  {1990})}\BibitemShut {NoStop}%
\bibitem [{\citenamefont {Souza}\ and\ \citenamefont
  {Lutz}(1977)}]{77SoLuxx.c2m}%
  \BibitemOpen
  \bibfield  {author} {\bibinfo {author} {\bibfnamefont {S.~P.}\ \bibnamefont
  {Souza}}\ and\ \bibinfo {author} {\bibfnamefont {B.~L.}\ \bibnamefont
  {Lutz}},\ }\href {\doibase 10.1086/182507} {\bibfield  {journal} {\bibinfo
  {journal} {Astrophys J.}\ }\textbf {\bibinfo {volume} {216}},\ \bibinfo
  {pages} {L49} (\bibinfo {year} {1977})}\BibitemShut {NoStop}%
\bibitem [{\citenamefont {Lambert}\ \emph {et~al.}(1986)\citenamefont
  {Lambert}, \citenamefont {Gustafsson}, \citenamefont {Eriksson},\ and\
  \citenamefont {Hinkle}}]{86LaGuEr.c2m}%
  \BibitemOpen
  \bibfield  {author} {\bibinfo {author} {\bibfnamefont {D.~L.}\ \bibnamefont
  {Lambert}}, \bibinfo {author} {\bibfnamefont {B.}~\bibnamefont {Gustafsson}},
  \bibinfo {author} {\bibfnamefont {K.}~\bibnamefont {Eriksson}}, \ and\
  \bibinfo {author} {\bibfnamefont {K.~H.}\ \bibnamefont {Hinkle}},\ }\href
  {\doibase 10.1086/191145} {\bibfield  {journal} {\bibinfo  {journal}
  {Astrophys J. Suppl. Ser.}\ }\textbf {\bibinfo {volume} {62}},\ \bibinfo
  {pages} {373} (\bibinfo {year} {1986})}\BibitemShut {NoStop}%
\bibitem [{\citenamefont {Vardya}\ and\ \citenamefont
  {Krishna~Swamy}(1980)}]{80VaSwxx.c2m}%
  \BibitemOpen
  \bibfield  {author} {\bibinfo {author} {\bibfnamefont {M.~S.}\ \bibnamefont
  {Vardya}}\ and\ \bibinfo {author} {\bibfnamefont {K.~S.}\ \bibnamefont
  {Krishna~Swamy}},\ }\href {\doibase 10.1016/0009-2614(80)80730-5} {\bibfield
  {journal} {\bibinfo  {journal} {Chem. Phys. Lett.}\ }\textbf {\bibinfo
  {volume} {73}},\ \bibinfo {pages} {616} (\bibinfo {year} {1980})}\BibitemShut
  {NoStop}%
\bibitem [{\citenamefont {Fa\"{y}}\ and\ \citenamefont
  {Johnson}(1972)}]{72FaJoxx.c2m}%
  \BibitemOpen
  \bibfield  {author} {\bibinfo {author} {\bibfnamefont {T.}~\bibnamefont
  {Fa\"{y}}}\ and\ \bibinfo {author} {\bibfnamefont {H.~R.}\ \bibnamefont
  {Johnson}},\ }\href {\doibase 10.1086/129284} {\bibfield  {journal} {\bibinfo
   {journal} {PASP}\ }\textbf {\bibinfo {volume} {84}},\ \bibinfo {pages} {284}
  (\bibinfo {year} {1972})}\BibitemShut {NoStop}%
\bibitem [{\citenamefont {Wallerstein}(1982)}]{82Waxxxx.c2m}%
  \BibitemOpen
  \bibfield  {author} {\bibinfo {author} {\bibfnamefont {G.}~\bibnamefont
  {Wallerstein}},\ }\href@noop {} {\bibfield  {journal} {\bibinfo  {journal}
  {Astron. Astrophys.}\ }\textbf {\bibinfo {volume} {105}},\ \bibinfo {pages}
  {219} (\bibinfo {year} {1982})}\BibitemShut {NoStop}%
\bibitem [{\citenamefont {Civi\u{s}}\ \emph {et~al.}(2005)\citenamefont
  {Civi\u{s}}, \citenamefont {Hosaki}, \citenamefont {Kagi}, \citenamefont
  {Izumiura}, \citenamefont {Yanagisawa}, \citenamefont {\u{S}edivcov\'{a}},\
  and\ \citenamefont {Kawaguchi}}]{05CiHoKa.c2m}%
  \BibitemOpen
  \bibfield  {author} {\bibinfo {author} {\bibfnamefont {S.}~\bibnamefont
  {Civi\u{s}}}, \bibinfo {author} {\bibfnamefont {Y.}~\bibnamefont {Hosaki}},
  \bibinfo {author} {\bibfnamefont {E.}~\bibnamefont {Kagi}}, \bibinfo {author}
  {\bibfnamefont {H.}~\bibnamefont {Izumiura}}, \bibinfo {author}
  {\bibfnamefont {K.}~\bibnamefont {Yanagisawa}}, \bibinfo {author}
  {\bibfnamefont {T.}~\bibnamefont {\u{S}edivcov\'{a}}}, \ and\ \bibinfo
  {author} {\bibfnamefont {K.}~\bibnamefont {Kawaguchi}},\ }\href {\doibase
  https://doi.org/10.1093/pasj/57.4.605} {\bibfield  {journal} {\bibinfo
  {journal} {Publ. Astron. Soc. Japan}\ }\textbf {\bibinfo {volume} {57}},\
  \bibinfo {pages} {605} (\bibinfo {year} {2005})}\BibitemShut {NoStop}%
\bibitem [{\citenamefont {Wester}(2009)}]{Wester2009:jpb}%
  \BibitemOpen
  \bibfield  {author} {\bibinfo {author} {\bibfnamefont {R.}~\bibnamefont
  {Wester}},\ }\href {\doibase 10.1088/0953-4075/42/15/154001} {\bibfield
  {journal} {\bibinfo  {journal} {J. Phys. B}\ }\textbf {\bibinfo {volume}
  {42}},\ \bibinfo {pages} {154001} (\bibinfo {year} {2009})}\BibitemShut
  {NoStop}%
\bibitem [{\citenamefont {Gianturco}\ \emph {et~al.}(2019)\citenamefont
  {Gianturco}, \citenamefont {Gonz\'{a}lez-S\'{a}nchez}, \citenamefont {Mant},\
  and\ \citenamefont {Wester}}]{19GiGoMa.c2hm}%
  \BibitemOpen
  \bibfield  {author} {\bibinfo {author} {\bibfnamefont {F.~A.}\ \bibnamefont
  {Gianturco}}, \bibinfo {author} {\bibfnamefont {L.}~\bibnamefont
  {Gonz\'{a}lez-S\'{a}nchez}}, \bibinfo {author} {\bibfnamefont {B.~P.}\
  \bibnamefont {Mant}}, \ and\ \bibinfo {author} {\bibfnamefont
  {R.}~\bibnamefont {Wester}},\ }\href {\doibase 10.1063/1.5123218} {\bibfield
  {journal} {\bibinfo  {journal} {J. Chem. Phys.}\ }\textbf {\bibinfo {volume}
  {151}},\ \bibinfo {pages} {144304} (\bibinfo {year} {2019})}\BibitemShut
  {NoStop}%
\bibitem [{\citenamefont {Hinterberger}\ \emph {et~al.}(2019)\citenamefont
  {Hinterberger}, \citenamefont {Gerber}, \citenamefont {Oswald}, \citenamefont
  {Zimmer}, \citenamefont {Fesel},\ and\ \citenamefont {Doser}}]{19HiGeOs}%
  \BibitemOpen
  \bibfield  {author} {\bibinfo {author} {\bibfnamefont {A.}~\bibnamefont
  {Hinterberger}}, \bibinfo {author} {\bibfnamefont {S.}~\bibnamefont
  {Gerber}}, \bibinfo {author} {\bibfnamefont {E.}~\bibnamefont {Oswald}},
  \bibinfo {author} {\bibfnamefont {C.}~\bibnamefont {Zimmer}}, \bibinfo
  {author} {\bibfnamefont {J.}~\bibnamefont {Fesel}}, \ and\ \bibinfo {author}
  {\bibfnamefont {M.}~\bibnamefont {Doser}},\ }\href {\doibase
  10.1088/1361-6455/ab4940} {\bibfield  {journal} {\bibinfo  {journal} {J.
  Phys. B: At. Mol. Opt. Phys.}\ }\textbf {\bibinfo {volume} {52}},\ \bibinfo
  {pages} {225003} (\bibinfo {year} {2019})}\BibitemShut {NoStop}%
\bibitem [{\citenamefont {Hansen}\ \emph {et~al.}(2014)\citenamefont {Hansen},
  \citenamefont {Versolato}, \citenamefont {K\l{}osowski}, \citenamefont
  {Kristensen}, \citenamefont {Gingell}, \citenamefont {Schwarz}, \citenamefont
  {Windberer}, \citenamefont {Ullrich}, \citenamefont {Crespo
  L\'{o}pez-Urrutia},\ and\ \citenamefont {Drewsen}}]{14HaVeKl}%
  \BibitemOpen
  \bibfield  {author} {\bibinfo {author} {\bibfnamefont {A.~K.}\ \bibnamefont
  {Hansen}}, \bibinfo {author} {\bibfnamefont {O.~O.}\ \bibnamefont
  {Versolato}}, \bibinfo {author} {\bibfnamefont {L.}~\bibnamefont
  {K\l{}osowski}}, \bibinfo {author} {\bibfnamefont {S.~B.}\ \bibnamefont
  {Kristensen}}, \bibinfo {author} {\bibfnamefont {A.}~\bibnamefont {Gingell}},
  \bibinfo {author} {\bibfnamefont {M.}~\bibnamefont {Schwarz}}, \bibinfo
  {author} {\bibfnamefont {K.}~\bibnamefont {Windberer}}, \bibinfo {author}
  {\bibfnamefont {K.}~\bibnamefont {Ullrich}}, \bibinfo {author} {\bibfnamefont
  {J.~R.}\ \bibnamefont {Crespo L\'{o}pez-Urrutia}}, \ and\ \bibinfo {author}
  {\bibfnamefont {M.}~\bibnamefont {Drewsen}},\ }\href {\doibase
  10.1038/nature12996} {\bibfield  {journal} {\bibinfo  {journal} {Nature}\
  }\textbf {\bibinfo {volume} {508}},\ \bibinfo {pages} {76} (\bibinfo {year}
  {2014})}\BibitemShut {NoStop}%
\bibitem [{\citenamefont {Mant}\ \emph
  {et~al.}(2020{\natexlab{a}})\citenamefont {Mant}, \citenamefont {Gianturco},
  \citenamefont {Gonz\'alez-S\'anchez}, \citenamefont {Yurtsever},\ and\
  \citenamefont {Wester}}]{20MaGiGo}%
  \BibitemOpen
  \bibfield  {author} {\bibinfo {author} {\bibfnamefont {B.~P.}\ \bibnamefont
  {Mant}}, \bibinfo {author} {\bibfnamefont {F.~A.}\ \bibnamefont {Gianturco}},
  \bibinfo {author} {\bibfnamefont {L.}~\bibnamefont {Gonz\'alez-S\'anchez}},
  \bibinfo {author} {\bibfnamefont {E.}~\bibnamefont {Yurtsever}}, \ and\
  \bibinfo {author} {\bibfnamefont {R.}~\bibnamefont {Wester}},\ }\href
  {\doibase 10.1088/1361-6455/ab574f} {\bibfield  {journal} {\bibinfo
  {journal} {J. Phys. B: At. Mol. Opt. Phys.}\ }\textbf {\bibinfo {volume}
  {53}},\ \bibinfo {pages} {025201} (\bibinfo {year}
  {2020}{\natexlab{a}})}\BibitemShut {NoStop}%
\bibitem [{\citenamefont {Mant}\ \emph
  {et~al.}(2020{\natexlab{b}})\citenamefont {Mant}, \citenamefont {Gianturco},
  \citenamefont {Wester}, \citenamefont {Yurtsever},\ and\ \citenamefont
  {Gonz\'{a}lez-S\'{a}nchez}}]{20aMaGiWe}%
  \BibitemOpen
  \bibfield  {author} {\bibinfo {author} {\bibfnamefont {B.~P.}\ \bibnamefont
  {Mant}}, \bibinfo {author} {\bibfnamefont {F.~A.}\ \bibnamefont {Gianturco}},
  \bibinfo {author} {\bibfnamefont {R.}~\bibnamefont {Wester}}, \bibinfo
  {author} {\bibfnamefont {E.}~\bibnamefont {Yurtsever}}, \ and\ \bibinfo
  {author} {\bibfnamefont {L.}~\bibnamefont {Gonz\'{a}lez-S\'{a}nchez}},\
  }\href {\doibase 10.1016/j.ijms.2020.116426} {\bibfield  {journal} {\bibinfo
  {journal} {J. Int. Mass Spectrom.}\ }\textbf {\bibinfo {volume} {457}},\
  \bibinfo {pages} {116426} (\bibinfo {year} {2020}{\natexlab{b}})}\BibitemShut
  {NoStop}%
\bibitem [{\citenamefont {Le~Roy}(2017{\natexlab{a}})}]{17RKR1}%
  \BibitemOpen
  \bibfield  {author} {\bibinfo {author} {\bibfnamefont {R.~J.}\ \bibnamefont
  {Le~Roy}},\ }\href {\doibase 10.1016/j.jqsrt.2016.03.030} {\bibfield
  {journal} {\bibinfo  {journal} {J. Quant. Spectrosc. Radiat. Transf.}\
  }\textbf {\bibinfo {volume} {186}},\ \bibinfo {pages} {158} (\bibinfo {year}
  {2017}{\natexlab{a}})}\BibitemShut {NoStop}%
\bibitem [{\citenamefont {Le~Roy}(2017{\natexlab{b}})}]{17LEVEL}%
  \BibitemOpen
  \bibfield  {author} {\bibinfo {author} {\bibfnamefont {R.~J.}\ \bibnamefont
  {Le~Roy}},\ }\href {\doibase 10.1016/j.jqsrt.2016.05.028} {\bibfield
  {journal} {\bibinfo  {journal} {J. Quant. Spectrosc. Radiat. Transf.}\
  }\textbf {\bibinfo {volume} {186}},\ \bibinfo {pages} {167} (\bibinfo {year}
  {2017}{\natexlab{b}})}\BibitemShut {NoStop}%
\bibitem [{pra()}]{pra_supp}%
  \BibitemOpen
  \href@noop {} {\enquote {\bibinfo {title} {See supplemental material at [url
  will be inserted by publisher] for details of the C$_2^-$ PEC fit and
  vibrational wavefunctions, Fortran programs for the 2D C$_2^-$-He/Ne/Ar rigid
  rotor PES and Legendre expansion parameters, Fortran programs for the 3D
  C$_2^-$-He/Ne/Ar interaction potential and Legendre expansions of the matrix
  elements of these surfaces},}\ }\BibitemShut {NoStop}%
\bibitem [{\citenamefont {Bunker}(1974)}]{74Buxxxx}%
  \BibitemOpen
  \bibfield  {author} {\bibinfo {author} {\bibfnamefont {P.~R.}\ \bibnamefont
  {Bunker}},\ }\href {\doibase 10.1016/0009-2614(74)90233-4} {\bibfield
  {journal} {\bibinfo  {journal} {Chem. Phys. Lett.}\ }\textbf {\bibinfo
  {volume} {27}},\ \bibinfo {pages} {322} (\bibinfo {year} {1974})}\BibitemShut
  {NoStop}%
\bibitem [{\citenamefont {Ellison}(1962)}]{62Elxxxx}%
  \BibitemOpen
  \bibfield  {author} {\bibinfo {author} {\bibfnamefont {F.~O.}\ \bibnamefont
  {Ellison}},\ }\href {\doibase 10.1063/1.1732535} {\bibfield  {journal}
  {\bibinfo  {journal} {J. Chem. Phys.}\ }\textbf {\bibinfo {volume} {36}},\
  \bibinfo {pages} {478} (\bibinfo {year} {1962})}\BibitemShut {NoStop}%
\bibitem [{\citenamefont {Ishikawa}\ \emph {et~al.}(2012)\citenamefont
  {Ishikawa}, \citenamefont {Nakashima},\ and\ \citenamefont
  {Nakatsuji}}]{12IsNaNa}%
  \BibitemOpen
  \bibfield  {author} {\bibinfo {author} {\bibfnamefont {A.}~\bibnamefont
  {Ishikawa}}, \bibinfo {author} {\bibfnamefont {H.}~\bibnamefont {Nakashima}},
  \ and\ \bibinfo {author} {\bibfnamefont {H.}~\bibnamefont {Nakatsuji}},\
  }\href {\doibase 10.1016/j.chemphys.2011.09.013} {\bibfield  {journal}
  {\bibinfo  {journal} {Chemical Physics}\ }\textbf {\bibinfo {volume} {401}},\
  \bibinfo {pages} {62} (\bibinfo {year} {2012})}\BibitemShut {NoStop}%
\bibitem [{\citenamefont {Dumouchel}\ \emph {et~al.}(2012)\citenamefont
  {Dumouchel}, \citenamefont {Spielfiedel}, \citenamefont {Senent},\ and\
  \citenamefont {Feautrier}}]{12DuSpSe}%
  \BibitemOpen
  \bibfield  {author} {\bibinfo {author} {\bibfnamefont {F.}~\bibnamefont
  {Dumouchel}}, \bibinfo {author} {\bibfnamefont {A.}~\bibnamefont
  {Spielfiedel}}, \bibinfo {author} {\bibfnamefont {M.}~\bibnamefont {Senent}},
  \ and\ \bibinfo {author} {\bibfnamefont {N.}~\bibnamefont {Feautrier}},\
  }\href {\doibase https://doi.org/10.1016/j.cplett.2012.03.006} {\bibfield
  {journal} {\bibinfo  {journal} {Chemical Physics Letters}\ }\textbf {\bibinfo
  {volume} {533}},\ \bibinfo {pages} {6} (\bibinfo {year} {2012})}\BibitemShut
  {NoStop}%
\bibitem [{\citenamefont {Br\"{u}nken}\ \emph {et~al.}(2007)\citenamefont
  {Br\"{u}nken}, \citenamefont {Gottlieb}, \citenamefont {Gupta}, \citenamefont
  {McCarthy},\ and\ \citenamefont {Thaddeus}}]{07BrGoGu}%
  \BibitemOpen
  \bibfield  {author} {\bibinfo {author} {\bibfnamefont {S.}~\bibnamefont
  {Br\"{u}nken}}, \bibinfo {author} {\bibfnamefont {C.~A.}\ \bibnamefont
  {Gottlieb}}, \bibinfo {author} {\bibfnamefont {H.}~\bibnamefont {Gupta}},
  \bibinfo {author} {\bibfnamefont {M.~C.}\ \bibnamefont {McCarthy}}, \ and\
  \bibinfo {author} {\bibfnamefont {P.}~\bibnamefont {Thaddeus}},\ }\href
  {\doibase 10.1051/0004-6361:20066964} {\bibfield  {journal} {\bibinfo
  {journal} {A\&A}\ }\textbf {\bibinfo {volume} {464}},\ \bibinfo {pages} {L33}
  (\bibinfo {year} {2007})}\BibitemShut {NoStop}%
\bibitem [{\citenamefont {Werner}\ \emph {et~al.}(2012)\citenamefont {Werner},
  \citenamefont {Knowles}, \citenamefont {Knizia}, \citenamefont {Manby},\ and\
  \citenamefont {Sch\"utz}}]{MOLPRO}%
  \BibitemOpen
  \bibfield  {author} {\bibinfo {author} {\bibfnamefont {H.-J.}\ \bibnamefont
  {Werner}}, \bibinfo {author} {\bibfnamefont {P.~J.}\ \bibnamefont {Knowles}},
  \bibinfo {author} {\bibfnamefont {G.}~\bibnamefont {Knizia}}, \bibinfo
  {author} {\bibfnamefont {F.~R.}\ \bibnamefont {Manby}}, \ and\ \bibinfo
  {author} {\bibfnamefont {M.}~\bibnamefont {Sch\"utz}},\ }\href {\doibase
  10.1002/wcms.82} {\bibfield  {journal} {\bibinfo  {journal} {WIREs Comput.
  Mol. Sci.}\ }\textbf {\bibinfo {volume} {2}},\ \bibinfo {pages} {242}
  (\bibinfo {year} {2012})}\BibitemShut {NoStop}%
\bibitem [{\citenamefont {Werner}\ \emph {et~al.}(2019)\citenamefont {Werner},
  \citenamefont {Knowles}, \citenamefont {Knizia}, \citenamefont {Manby},
  \citenamefont {{Sch\"{u}tz}} \emph {et~al.}}]{MOLPRO_brief}%
  \BibitemOpen
  \bibfield  {author} {\bibinfo {author} {\bibfnamefont {H.-J.}\ \bibnamefont
  {Werner}}, \bibinfo {author} {\bibfnamefont {P.~J.}\ \bibnamefont {Knowles}},
  \bibinfo {author} {\bibfnamefont {G.}~\bibnamefont {Knizia}}, \bibinfo
  {author} {\bibfnamefont {F.~R.}\ \bibnamefont {Manby}}, \bibinfo {author}
  {\bibfnamefont {M.}~\bibnamefont {{Sch\"{u}tz}}},  \emph {et~al.},\
  }\href@noop {} {\enquote {\bibinfo {title} {Molpro, version 2019.2, a package
  of ab initio programs},}\ } (\bibinfo {year} {2019}),\ \bibinfo {note} {see
  https://www.molpro.net}\BibitemShut {NoStop}%
\bibitem [{\citenamefont {Werner}\ and\ \citenamefont
  {Knowles}(1985)}]{85WeKnxx}%
  \BibitemOpen
  \bibfield  {author} {\bibinfo {author} {\bibfnamefont {H.~J.}\ \bibnamefont
  {Werner}}\ and\ \bibinfo {author} {\bibfnamefont {P.~J.}\ \bibnamefont
  {Knowles}},\ }\href {\doibase 10.1063/1.448627} {\bibfield  {journal}
  {\bibinfo  {journal} {J. Chem. Phys.}\ }\textbf {\bibinfo {volume} {82}},\
  \bibinfo {pages} {5053} (\bibinfo {year} {1985})}\BibitemShut {NoStop}%
\bibitem [{\citenamefont {Knowles}\ and\ \citenamefont
  {Werner}(1985)}]{85KnWexx}%
  \BibitemOpen
  \bibfield  {author} {\bibinfo {author} {\bibfnamefont {P.~J.}\ \bibnamefont
  {Knowles}}\ and\ \bibinfo {author} {\bibfnamefont {H.~J.}\ \bibnamefont
  {Werner}},\ }\href {\doibase 10.1016/0009-2614(85)80025-7} {\bibfield
  {journal} {\bibinfo  {journal} {Chem. Phys. Lett.}\ }\textbf {\bibinfo
  {volume} {115}},\ \bibinfo {pages} {259} (\bibinfo {year}
  {1985})}\BibitemShut {NoStop}%
\bibitem [{\citenamefont {Shamasundar}\ \emph {et~al.}(2011)\citenamefont
  {Shamasundar}, \citenamefont {Knizia},\ and\ \citenamefont
  {Werner}}]{11ShKnWe.LM}%
  \BibitemOpen
  \bibfield  {author} {\bibinfo {author} {\bibfnamefont {K.~R.}\ \bibnamefont
  {Shamasundar}}, \bibinfo {author} {\bibfnamefont {G.}~\bibnamefont {Knizia}},
  \ and\ \bibinfo {author} {\bibfnamefont {H.-J.}\ \bibnamefont {Werner}},\
  }\href {\doibase 10.1063/1.3609809} {\bibfield  {journal} {\bibinfo
  {journal} {J. Chem. Phys.}\ }\textbf {\bibinfo {volume} {135}},\ \bibinfo
  {pages} {053101} (\bibinfo {year} {2011})}\BibitemShut {NoStop}%
\bibitem [{\citenamefont {Kendall}\ \emph {et~al.}(1992)\citenamefont
  {Kendall}, \citenamefont {Dunning~Jr},\ and\ \citenamefont
  {Harrison}}]{92KeDuHa}%
  \BibitemOpen
  \bibfield  {author} {\bibinfo {author} {\bibfnamefont {R.~A.}\ \bibnamefont
  {Kendall}}, \bibinfo {author} {\bibfnamefont {T.~H.}\ \bibnamefont
  {Dunning~Jr}}, \ and\ \bibinfo {author} {\bibfnamefont {R.~J.}\ \bibnamefont
  {Harrison}},\ }\href {\doibase 10.1063/1.462569} {\bibfield  {journal}
  {\bibinfo  {journal} {J. Chem. Phys.}\ }\textbf {\bibinfo {volume} {96}},\
  \bibinfo {pages} {6796} (\bibinfo {year} {1992})}\BibitemShut {NoStop}%
\bibitem [{\citenamefont {Knowles}\ \emph {et~al.}(1993)\citenamefont
  {Knowles}, \citenamefont {Hampel},\ and\ \citenamefont {Werner}}]{93KnHaWe}%
  \BibitemOpen
  \bibfield  {author} {\bibinfo {author} {\bibfnamefont {P.~J.}\ \bibnamefont
  {Knowles}}, \bibinfo {author} {\bibfnamefont {C.}~\bibnamefont {Hampel}}, \
  and\ \bibinfo {author} {\bibfnamefont {H.-J.}\ \bibnamefont {Werner}},\
  }\href {\doibase 10.1063/1.465990} {\bibfield  {journal} {\bibinfo  {journal}
  {J. Chem. Phys.}\ }\textbf {\bibinfo {volume} {99}},\ \bibinfo {pages} {5219}
  (\bibinfo {year} {1993})}\BibitemShut {NoStop}%
\bibitem [{\citenamefont {Deegan}\ and\ \citenamefont
  {Knowles}(1994)}]{94DeKnxx}%
  \BibitemOpen
  \bibfield  {author} {\bibinfo {author} {\bibfnamefont {M.~J.~O.}\
  \bibnamefont {Deegan}}\ and\ \bibinfo {author} {\bibfnamefont {P.~J.}\
  \bibnamefont {Knowles}},\ }\href {\doibase 10.1016/0009-2614(94)00815-9}
  {\bibfield  {journal} {\bibinfo  {journal} {Chem. Phys. Lett.}\ }\textbf
  {\bibinfo {volume} {227}},\ \bibinfo {pages} {321} (\bibinfo {year}
  {1994})}\BibitemShut {NoStop}%
\bibitem [{\citenamefont {Wilson}\ and\ \citenamefont {van
  Mourik}(1996)}]{96WiMoDu}%
  \BibitemOpen
  \bibfield  {author} {\bibinfo {author} {\bibfnamefont {A.~K.}\ \bibnamefont
  {Wilson}}\ and\ \bibinfo {author} {\bibfnamefont {T.~H.}\ \bibnamefont {van
  Mourik}, \bibfnamefont {T~amd~Dunning}},\ }\href {\doibase
  10.1016/S0166-1280(96)80048-0} {\bibfield  {journal} {\bibinfo  {journal}
  {Theochem}\ }\textbf {\bibinfo {volume} {388}},\ \bibinfo {pages} {339}
  (\bibinfo {year} {1996})}\BibitemShut {NoStop}%
\bibitem [{\citenamefont {Woon}\ and\ \citenamefont
  {Dunning~Jr}(1993)}]{93WoDuxx}%
  \BibitemOpen
  \bibfield  {author} {\bibinfo {author} {\bibfnamefont {D.~E.}\ \bibnamefont
  {Woon}}\ and\ \bibinfo {author} {\bibfnamefont {T.~H.}\ \bibnamefont
  {Dunning~Jr}},\ }\href {\doibase 10.1063/1.464303} {\bibfield  {journal}
  {\bibinfo  {journal} {J. Chem. Phys.}\ }\textbf {\bibinfo {volume} {98}},\
  \bibinfo {pages} {1358} (\bibinfo {year} {1993})}\BibitemShut {NoStop}%
\bibitem [{\citenamefont {Boys}\ and\ \citenamefont
  {Bernardi}(1970)}]{70BoBexx}%
  \BibitemOpen
  \bibfield  {author} {\bibinfo {author} {\bibfnamefont {S.~F.}\ \bibnamefont
  {Boys}}\ and\ \bibinfo {author} {\bibfnamefont {F.}~\bibnamefont
  {Bernardi}},\ }\href {\doibase 10.1080/00268977000101561} {\bibfield
  {journal} {\bibinfo  {journal} {Mol. Phys.}\ }\textbf {\bibinfo {volume}
  {19}},\ \bibinfo {pages} {553} (\bibinfo {year} {1970})}\BibitemShut
  {NoStop}%
\bibitem [{\citenamefont {Werner}\ \emph {et~al.}(1988)\citenamefont {Werner},
  \citenamefont {Follmeg},\ and\ \citenamefont {Alexander}}]{88WeFoAl}%
  \BibitemOpen
  \bibfield  {author} {\bibinfo {author} {\bibfnamefont {H.-J.}\ \bibnamefont
  {Werner}}, \bibinfo {author} {\bibfnamefont {B.}~\bibnamefont {Follmeg}}, \
  and\ \bibinfo {author} {\bibfnamefont {M.}~\bibnamefont {Alexander}},\ }\href
  {\doibase 10.1063/1.454971} {\bibfield  {journal} {\bibinfo  {journal} {J.
  Chem. Phys.}\ }\textbf {\bibinfo {volume} {89}},\ \bibinfo {pages} {3139}
  (\bibinfo {year} {1988})}\BibitemShut {NoStop}%
\bibitem [{\citenamefont {Balan\c{c}a}\ and\ \citenamefont
  {Dayou}(2017)}]{17BaDaxx}%
  \BibitemOpen
  \bibfield  {author} {\bibinfo {author} {\bibfnamefont {C.}~\bibnamefont
  {Balan\c{c}a}}\ and\ \bibinfo {author} {\bibfnamefont {F.}~\bibnamefont
  {Dayou}},\ }\href {\doibase 10.1093/mnras/stx925} {\bibfield  {journal}
  {\bibinfo  {journal} {MNRAS}\ }\textbf {\bibinfo {volume} {469}},\ \bibinfo
  {pages} {1673} (\bibinfo {year} {2017})}\BibitemShut {NoStop}%
\bibitem [{\citenamefont {Gaiser}\ and\ \citenamefont
  {Fellmuth}(2018)}]{18GaFexx}%
  \BibitemOpen
  \bibfield  {author} {\bibinfo {author} {\bibfnamefont {C.}~\bibnamefont
  {Gaiser}}\ and\ \bibinfo {author} {\bibfnamefont {B.}~\bibnamefont
  {Fellmuth}},\ }\href {\doibase 10.1103/PhysRevLett.120.123203} {\bibfield
  {journal} {\bibinfo  {journal} {Phys. Rev. Lett.}\ }\textbf {\bibinfo
  {volume} {120}},\ \bibinfo {pages} {123203} (\bibinfo {year}
  {2018})}\BibitemShut {NoStop}%
\bibitem [{\citenamefont {L\'opez-Dur\'ann}\ \emph {et~al.}(2008)\citenamefont
  {L\'opez-Dur\'ann}, \citenamefont {Bodo},\ and\ \citenamefont
  {Gianturco}}]{08LoBoGi}%
  \BibitemOpen
  \bibfield  {author} {\bibinfo {author} {\bibfnamefont {D.}~\bibnamefont
  {L\'opez-Dur\'ann}}, \bibinfo {author} {\bibfnamefont {E.}~\bibnamefont
  {Bodo}}, \ and\ \bibinfo {author} {\bibfnamefont {F.~A.}\ \bibnamefont
  {Gianturco}},\ }\href {\doibase 10.1016/j.cpc.2008.07.017} {\bibfield
  {journal} {\bibinfo  {journal} {Comput. Phys. Commun.}\ }\textbf {\bibinfo
  {volume} {179}},\ \bibinfo {pages} {821} (\bibinfo {year}
  {2008})}\BibitemShut {NoStop}%
\bibitem [{\citenamefont {Arthurs}\ and\ \citenamefont
  {Dalgarno}(1960)}]{60ArDaxx}%
  \BibitemOpen
  \bibfield  {author} {\bibinfo {author} {\bibfnamefont {A.~M.}\ \bibnamefont
  {Arthurs}}\ and\ \bibinfo {author} {\bibfnamefont {A.}~\bibnamefont
  {Dalgarno}},\ }\href {\doibase 10.1098/rspa.1960.0125} {\bibfield  {journal}
  {\bibinfo  {journal} {Proc. R. Soc. A}\ }\textbf {\bibinfo {volume} {256}},\
  \bibinfo {pages} {540} (\bibinfo {year} {1960})}\BibitemShut {NoStop}%
\bibitem [{\citenamefont {Manolopoulos}(1986)}]{86Maxxxx.c2m}%
  \BibitemOpen
  \bibfield  {author} {\bibinfo {author} {\bibfnamefont {D.~E.}\ \bibnamefont
  {Manolopoulos}},\ }\href {\doibase 10.1063/1.451472} {\bibfield  {journal}
  {\bibinfo  {journal} {J. Chem. Phys.}\ }\textbf {\bibinfo {volume} {85}},\
  \bibinfo {pages} {6425} (\bibinfo {year} {1986})}\BibitemShut {NoStop}%
\bibitem [{\citenamefont {Martinazzo}\ \emph {et~al.}(2003)\citenamefont
  {Martinazzo}, \citenamefont {Bodo},\ and\ \citenamefont
  {Gianturco}}]{03MaBoGi}%
  \BibitemOpen
  \bibfield  {author} {\bibinfo {author} {\bibfnamefont {R.}~\bibnamefont
  {Martinazzo}}, \bibinfo {author} {\bibfnamefont {E.}~\bibnamefont {Bodo}}, \
  and\ \bibinfo {author} {\bibfnamefont {F.~A.}\ \bibnamefont {Gianturco}},\
  }\href {\doibase 10.1016/S0010-4655(02)00737-3} {\bibfield  {journal}
  {\bibinfo  {journal} {Comput. Phys. Commun.}\ }\textbf {\bibinfo {volume}
  {151}},\ \bibinfo {pages} {187} (\bibinfo {year} {2003})}\BibitemShut
  {NoStop}%
\bibitem [{\citenamefont {Iskandarov}\ \emph {et~al.}(2017)\citenamefont
  {Iskandarov}, \citenamefont {Gianturco}, \citenamefont {Hern\'andez~Vera},
  \citenamefont {Wester}, \citenamefont {da~Silva~Jr.},\ and\ \citenamefont
  {Dulieu}}]{17IsGiHe}%
  \BibitemOpen
  \bibfield  {author} {\bibinfo {author} {\bibfnamefont {I.}~\bibnamefont
  {Iskandarov}}, \bibinfo {author} {\bibfnamefont {F.~A.}\ \bibnamefont
  {Gianturco}}, \bibinfo {author} {\bibfnamefont {M.}~\bibnamefont
  {Hern\'andez~Vera}}, \bibinfo {author} {\bibfnamefont {R.}~\bibnamefont
  {Wester}}, \bibinfo {author} {\bibfnamefont {H.}~\bibnamefont
  {da~Silva~Jr.}}, \ and\ \bibinfo {author} {\bibfnamefont {O.}~\bibnamefont
  {Dulieu}},\ }\href {\doibase 10.1140/epjd/e2017-80043-8} {\bibfield
  {journal} {\bibinfo  {journal} {Eur. Phys. J. D}\ }\textbf {\bibinfo {volume}
  {71}},\ \bibinfo {pages} {141} (\bibinfo {year} {2017})}\BibitemShut
  {NoStop}%
\bibitem [{\citenamefont {Krems}(2002)}]{02Krxxxx}%
  \BibitemOpen
  \bibfield  {author} {\bibinfo {author} {\bibfnamefont {R.~V.}\ \bibnamefont
  {Krems}},\ }\href {\doibase 10.1063/1.1451061} {\bibfield  {journal}
  {\bibinfo  {journal} {J. Chem. Phys.}\ }\textbf {\bibinfo {volume} {116}},\
  \bibinfo {pages} {4517} (\bibinfo {year} {2002})}\BibitemShut {NoStop}%
\bibitem [{\citenamefont {Lique}\ \emph {et~al.}(2006)\citenamefont {Lique},
  \citenamefont {Spielfiedel}, \citenamefont {Dhont},\ and\ \citenamefont
  {Feautrier}}]{06FiSpDh}%
  \BibitemOpen
  \bibfield  {author} {\bibinfo {author} {\bibfnamefont {F.}~\bibnamefont
  {Lique}}, \bibinfo {author} {\bibfnamefont {A.}~\bibnamefont {Spielfiedel}},
  \bibinfo {author} {\bibfnamefont {G.}~\bibnamefont {Dhont}}, \ and\ \bibinfo
  {author} {\bibfnamefont {N.}~\bibnamefont {Feautrier}},\ }\href {\doibase
  10.1051/0004-6361:20065713} {\bibfield  {journal} {\bibinfo  {journal}
  {Astron. Astrophys.}\ }\textbf {\bibinfo {volume} {458}},\ \bibinfo {pages}
  {331} (\bibinfo {year} {2006})}\BibitemShut {NoStop}%
\bibitem [{\citenamefont {Lique}\ and\ \citenamefont
  {Spielfiedel}(2007)}]{07FiSpxx}%
  \BibitemOpen
  \bibfield  {author} {\bibinfo {author} {\bibfnamefont {F.}~\bibnamefont
  {Lique}}\ and\ \bibinfo {author} {\bibfnamefont {A.}~\bibnamefont
  {Spielfiedel}},\ }\href {\doibase 10.1051/0004-6361:20066422} {\bibfield
  {journal} {\bibinfo  {journal} {Astron. Astrophys.}\ }\textbf {\bibinfo
  {volume} {462}},\ \bibinfo {pages} {1179} (\bibinfo {year}
  {2007})}\BibitemShut {NoStop}%
\bibitem [{\citenamefont {Tobo\l{}a}\ \emph {et~al.}(2008)\citenamefont
  {Tobo\l{}a}, \citenamefont {Lique}, \citenamefont {K\l{}os},\ and\
  \citenamefont {Cha\l{}asi\'{n}ski}}]{08ToLiKl}%
  \BibitemOpen
  \bibfield  {author} {\bibinfo {author} {\bibfnamefont {R.}~\bibnamefont
  {Tobo\l{}a}}, \bibinfo {author} {\bibfnamefont {F.}~\bibnamefont {Lique}},
  \bibinfo {author} {\bibfnamefont {J.}~\bibnamefont {K\l{}os}}, \ and\
  \bibinfo {author} {\bibfnamefont {G.}~\bibnamefont {Cha\l{}asi\'{n}ski}},\
  }\href {\doibase 10.1088/0953-4075/41/15/155702} {\bibfield  {journal}
  {\bibinfo  {journal} {J. Phys. B: At. Mol. Opt. Phys.}\ }\textbf {\bibinfo
  {volume} {41}},\ \bibinfo {pages} {155702} (\bibinfo {year}
  {2008})}\BibitemShut {NoStop}%
\bibitem [{\citenamefont {Krems}\ and\ \citenamefont
  {Nordholm}(2001)}]{01KrStxx}%
  \BibitemOpen
  \bibfield  {author} {\bibinfo {author} {\bibfnamefont {R.~V.}\ \bibnamefont
  {Krems}}\ and\ \bibinfo {author} {\bibfnamefont {S.}~\bibnamefont
  {Nordholm}},\ }\href {\doibase 10.1063/1.1378815} {\bibfield  {journal}
  {\bibinfo  {journal} {J. Chem. Phys.}\ }\textbf {\bibinfo {volume} {115}},\
  \bibinfo {pages} {257} (\bibinfo {year} {2001})}\BibitemShut {NoStop}%
\bibitem [{\citenamefont {Krems}\ \emph {et~al.}(2001)\citenamefont {Krems},
  \citenamefont {Markovi\'c}, \citenamefont {Buchachenko},\ and\ \citenamefont
  {Nordholm}}]{01KrMaBu}%
  \BibitemOpen
  \bibfield  {author} {\bibinfo {author} {\bibfnamefont {R.~V.}\ \bibnamefont
  {Krems}}, \bibinfo {author} {\bibfnamefont {N.}~\bibnamefont {Markovi\'c}},
  \bibinfo {author} {\bibfnamefont {A.~A.}\ \bibnamefont {Buchachenko}}, \ and\
  \bibinfo {author} {\bibfnamefont {S.}~\bibnamefont {Nordholm}},\ }\href
  {\doibase 10.1063/1.1333704} {\bibfield  {journal} {\bibinfo  {journal} {J.
  Chem. Phys.}\ }\textbf {\bibinfo {volume} {114}},\ \bibinfo {pages} {1249}
  (\bibinfo {year} {2001})}\BibitemShut {NoStop}%
\bibitem [{\citenamefont {Naja}\ \emph {et~al.}(2008)\citenamefont {Naja},
  \citenamefont {Ben~Abdallah}, \citenamefont {Jaidan},\ and\ \citenamefont
  {Lakhdar}}]{08NaAbJa}%
  \BibitemOpen
  \bibfield  {author} {\bibinfo {author} {\bibfnamefont {F.}~\bibnamefont
  {Naja}}, \bibinfo {author} {\bibfnamefont {D.}~\bibnamefont {Ben~Abdallah}},
  \bibinfo {author} {\bibfnamefont {N.}~\bibnamefont {Jaidan}}, \ and\ \bibinfo
  {author} {\bibfnamefont {B.}~\bibnamefont {Lakhdar}},\ }\href {\doibase
  10.1016/j.cplett.2008.05.070} {\bibfield  {journal} {\bibinfo  {journal}
  {Chem. Phys. Lett.}\ }\textbf {\bibinfo {volume} {460}},\ \bibinfo {pages}
  {31} (\bibinfo {year} {2008})}\BibitemShut {NoStop}%
\bibitem [{\citenamefont {George~D.X.}\ and\ \citenamefont
  {Kumar}(2010)}]{10GeKuxx}%
  \BibitemOpen
  \bibfield  {author} {\bibinfo {author} {\bibfnamefont {F.}~\bibnamefont
  {George~D.X.}}\ and\ \bibinfo {author} {\bibfnamefont {S.}~\bibnamefont
  {Kumar}},\ }\href {\doibase 10.1016/j.chemphys.2010.05.012} {\bibfield
  {journal} {\bibinfo  {journal} {Chem. Phys.}\ }\textbf {\bibinfo {volume}
  {373}},\ \bibinfo {pages} {211} (\bibinfo {year} {2010})}\BibitemShut
  {NoStop}%
\bibitem [{\citenamefont {Anusuri}\ and\ \citenamefont
  {Kumar}(2016)}]{16AnKuxx}%
  \BibitemOpen
  \bibfield  {author} {\bibinfo {author} {\bibfnamefont {B.}~\bibnamefont
  {Anusuri}}\ and\ \bibinfo {author} {\bibfnamefont {S.}~\bibnamefont
  {Kumar}},\ }\href {\doibase 0.1007/s12039-015-1022-8} {\bibfield  {journal}
  {\bibinfo  {journal} {J. Chem. Sci.}\ }\textbf {\bibinfo {volume} {128}},\
  \bibinfo {pages} {287} (\bibinfo {year} {2016})}\BibitemShut {NoStop}%
\bibitem [{\citenamefont {Saheer}\ and\ \citenamefont
  {Kumar}(2016)}]{16SaKuxx}%
  \BibitemOpen
  \bibfield  {author} {\bibinfo {author} {\bibfnamefont {V.~C.}\ \bibnamefont
  {Saheer}}\ and\ \bibinfo {author} {\bibfnamefont {S.}~\bibnamefont {Kumar}},\
  }\href {\doibase 10.1063/1.4939674} {\bibfield  {journal} {\bibinfo
  {journal} {J. Chem. Phys.}\ }\textbf {\bibinfo {volume} {144}},\ \bibinfo
  {pages} {024307} (\bibinfo {year} {2016})}\BibitemShut {NoStop}%
\bibitem [{\citenamefont {Xavier}\ and\ \citenamefont
  {Kumar}(2011)}]{11XaKuxx}%
  \BibitemOpen
  \bibfield  {author} {\bibinfo {author} {\bibfnamefont {F.~G.~D.}\
  \bibnamefont {Xavier}}\ and\ \bibinfo {author} {\bibfnamefont
  {S.}~\bibnamefont {Kumar}},\ }\href {\doibase 10.1103/PhysRevA.83.042709}
  {\bibfield  {journal} {\bibinfo  {journal} {Phys. Rev. A}\ }\textbf {\bibinfo
  {volume} {83}},\ \bibinfo {pages} {042709} (\bibinfo {year}
  {2011})}\BibitemShut {NoStop}%
\bibitem [{\citenamefont {Kendrick}(2019)}]{19Kexxxx}%
  \BibitemOpen
  \bibfield  {author} {\bibinfo {author} {\bibfnamefont {B.~K.}\ \bibnamefont
  {Kendrick}},\ }\href {\doibase 10.1021/acs.jpca.9b07318} {\bibfield
  {journal} {\bibinfo  {journal} {J. Phys. Chem. A}\ }\textbf {\bibinfo
  {volume} {123}},\ \bibinfo {pages} {9919} (\bibinfo {year}
  {2019})}\BibitemShut {NoStop}%
\bibitem [{\citenamefont {Kendrick}(2018)}]{18Kexxxx}%
  \BibitemOpen
  \bibfield  {author} {\bibinfo {author} {\bibfnamefont {B.~K.}\ \bibnamefont
  {Kendrick}},\ }\href {\doibase 10.1063/1.5014989} {\bibfield  {journal}
  {\bibinfo  {journal} {J. Chem. Phys.}\ }\textbf {\bibinfo {volume} {148}},\
  \bibinfo {pages} {044116} (\bibinfo {year} {2018})}\BibitemShut {NoStop}%
\bibitem [{\citenamefont {Kendrick}\ \emph {et~al.}(2016)\citenamefont
  {Kendrick}, \citenamefont {Hazra},\ and\ \citenamefont
  {Balakrishnan}}]{16KeHaBa}%
  \BibitemOpen
  \bibfield  {author} {\bibinfo {author} {\bibfnamefont {B.~K.}\ \bibnamefont
  {Kendrick}}, \bibinfo {author} {\bibfnamefont {J.}~\bibnamefont {Hazra}}, \
  and\ \bibinfo {author} {\bibfnamefont {N.}~\bibnamefont {Balakrishnan}},\
  }\href {\doibase 10.1063/1.4966037} {\bibfield  {journal} {\bibinfo
  {journal} {J. Chem. Phys.}\ }\textbf {\bibinfo {volume} {145}},\ \bibinfo
  {pages} {164303} (\bibinfo {year} {2016})}\BibitemShut {NoStop}%
\bibitem [{\citenamefont {Suleimanov}\ and\ \citenamefont
  {Tscherbul}(2016)}]{16SuTsxx}%
  \BibitemOpen
  \bibfield  {author} {\bibinfo {author} {\bibfnamefont {Y.~V.}\ \bibnamefont
  {Suleimanov}}\ and\ \bibinfo {author} {\bibfnamefont {T.~V.}\ \bibnamefont
  {Tscherbul}},\ }\href {\doibase 10.1088/0953-4075/49/20/204002} {\bibfield
  {journal} {\bibinfo  {journal} {J. Phys. B: At. Mol. Opt. Phys.}\ }\textbf
  {\bibinfo {volume} {49}},\ \bibinfo {pages} {204002} (\bibinfo {year}
  {2016})}\BibitemShut {NoStop}%
\bibitem [{\citenamefont {Morita}\ \emph {et~al.}(2019)\citenamefont {Morita},
  \citenamefont {Krems},\ and\ \citenamefont {Tscherbul}}]{19MoKrTs}%
  \BibitemOpen
  \bibfield  {author} {\bibinfo {author} {\bibfnamefont {M.}~\bibnamefont
  {Morita}}, \bibinfo {author} {\bibfnamefont {R.~V.}\ \bibnamefont {Krems}}, \
  and\ \bibinfo {author} {\bibfnamefont {T.~V.}\ \bibnamefont {Tscherbul}},\
  }\href {\doibase 10.1103/PhysRevLett.123.013401} {\bibfield  {journal}
  {\bibinfo  {journal} {Phys. Rev. Lett.}\ }\textbf {\bibinfo {volume} {123}},\
  \bibinfo {pages} {013401} (\bibinfo {year} {2019})}\BibitemShut {NoStop}%
\bibitem [{\citenamefont {Kato}\ \emph {et~al.}(1995)\citenamefont {Kato},
  \citenamefont {Bierbaum},\ and\ \citenamefont {Leone}}]{95KaBiLe}%
  \BibitemOpen
  \bibfield  {author} {\bibinfo {author} {\bibfnamefont {S.}~\bibnamefont
  {Kato}}, \bibinfo {author} {\bibfnamefont {V.~M.}\ \bibnamefont {Bierbaum}},
  \ and\ \bibinfo {author} {\bibfnamefont {S.~R.}\ \bibnamefont {Leone}},\
  }\href {\doibase 10.1016/0168-117(95)04283-Q} {\bibfield  {journal} {\bibinfo
   {journal} {Int. J. Mass Spec. Ion Proc.}\ }\textbf {\bibinfo {volume}
  {149/150}},\ \bibinfo {pages} {469} (\bibinfo {year} {1995})}\BibitemShut
  {NoStop}%
\bibitem [{\citenamefont {Ferguson}(1986)}]{86Fexxxx}%
  \BibitemOpen
  \bibfield  {author} {\bibinfo {author} {\bibfnamefont {E.~E.}\ \bibnamefont
  {Ferguson}},\ }\href {\doibase 10.1021/j100277a008} {\bibfield  {journal}
  {\bibinfo  {journal} {J. Phys. Chem.}\ }\textbf {\bibinfo {volume} {90}},\
  \bibinfo {pages} {731} (\bibinfo {year} {1986})}\BibitemShut {NoStop}%
\bibitem [{\citenamefont {Saidani}\ \emph {et~al.}(2013)\citenamefont
  {Saidani}, \citenamefont {Kalugina}, \citenamefont {Gardez}, \citenamefont
  {Biennier}, \citenamefont {Georges},\ and\ \citenamefont {Lique}}]{13SaKaGa}%
  \BibitemOpen
  \bibfield  {author} {\bibinfo {author} {\bibfnamefont {G.}~\bibnamefont
  {Saidani}}, \bibinfo {author} {\bibfnamefont {Y.}~\bibnamefont {Kalugina}},
  \bibinfo {author} {\bibfnamefont {A.}~\bibnamefont {Gardez}}, \bibinfo
  {author} {\bibfnamefont {L.}~\bibnamefont {Biennier}}, \bibinfo {author}
  {\bibfnamefont {R.}~\bibnamefont {Georges}}, \ and\ \bibinfo {author}
  {\bibfnamefont {F.}~\bibnamefont {Lique}},\ }\href {\doibase
  10.1063/1.4795206} {\bibfield  {journal} {\bibinfo  {journal} {J. Chem.
  Phys.}\ }\textbf {\bibinfo {volume} {138}},\ \bibinfo {pages} {124308}
  (\bibinfo {year} {2013})}\BibitemShut {NoStop}%
\bibitem [{\citenamefont {Stoecklin}\ and\ \citenamefont
  {Voronin}(2011)}]{11StVoxx}%
  \BibitemOpen
  \bibfield  {author} {\bibinfo {author} {\bibfnamefont {T.}~\bibnamefont
  {Stoecklin}}\ and\ \bibinfo {author} {\bibfnamefont {A.}~\bibnamefont
  {Voronin}},\ }\href {\doibase 10.1063/1.3590917} {\bibfield  {journal}
  {\bibinfo  {journal} {J. Chem. Phys.}\ }\textbf {\bibinfo {volume} {134}},\
  \bibinfo {pages} {204312} (\bibinfo {year} {2011})}\BibitemShut {NoStop}%
\bibitem [{\citenamefont {Caruso}\ \emph {et~al.}(2012)\citenamefont {Caruso},
  \citenamefont {Tacconi}, \citenamefont {Gianturco},\ and\ \citenamefont
  {Yurtsever}}]{12CaTaGi}%
  \BibitemOpen
  \bibfield  {author} {\bibinfo {author} {\bibfnamefont {D.}~\bibnamefont
  {Caruso}}, \bibinfo {author} {\bibfnamefont {M.}~\bibnamefont {Tacconi}},
  \bibinfo {author} {\bibfnamefont {F.~A.}\ \bibnamefont {Gianturco}}, \ and\
  \bibinfo {author} {\bibfnamefont {E.}~\bibnamefont {Yurtsever}},\ }\href
  {\doibase 10.1007/s12039-011-0190-4} {\bibfield  {journal} {\bibinfo
  {journal} {J. Chem. Sci.}\ }\textbf {\bibinfo {volume} {124}},\ \bibinfo
  {pages} {93} (\bibinfo {year} {2012})}\BibitemShut {NoStop}%
\bibitem [{\citenamefont {Stoecklin}\ and\ \citenamefont
  {Voronin}(2008)}]{08StVoxx}%
  \BibitemOpen
  \bibfield  {author} {\bibinfo {author} {\bibfnamefont {T.}~\bibnamefont
  {Stoecklin}}\ and\ \bibinfo {author} {\bibfnamefont {A.}~\bibnamefont
  {Voronin}},\ }\href {\doibase 10.1140/epjd/e2007-00293-3} {\bibfield
  {journal} {\bibinfo  {journal} {Eur. Phys. J. D}\ }\textbf {\bibinfo {volume}
  {46}},\ \bibinfo {pages} {259} (\bibinfo {year} {2008})}\BibitemShut
  {NoStop}%
\bibitem [{\citenamefont {Stoecklin}\ \emph {et~al.}(2016)\citenamefont
  {Stoecklin}, \citenamefont {Halvick}, \citenamefont {Gannounim},
  \citenamefont {Hochlaf}, \citenamefont {Kotochigova},\ and\ \citenamefont
  {Hudson}}]{16StHaGa}%
  \BibitemOpen
  \bibfield  {author} {\bibinfo {author} {\bibfnamefont {T.}~\bibnamefont
  {Stoecklin}}, \bibinfo {author} {\bibfnamefont {P.}~\bibnamefont {Halvick}},
  \bibinfo {author} {\bibfnamefont {M.~A.}\ \bibnamefont {Gannounim}}, \bibinfo
  {author} {\bibfnamefont {M.}~\bibnamefont {Hochlaf}}, \bibinfo {author}
  {\bibfnamefont {S.}~\bibnamefont {Kotochigova}}, \ and\ \bibinfo {author}
  {\bibfnamefont {E.~R.}\ \bibnamefont {Hudson}},\ }\href {\doibase
  10.1038/ncomms11234} {\bibfield  {journal} {\bibinfo  {journal} {Nat.
  Commun.}\ }\textbf {\bibinfo {volume} {7}},\ \bibinfo {pages} {11234}
  (\bibinfo {year} {2016})}\BibitemShut {NoStop}%
\bibitem [{\citenamefont {Campbell}\ \emph {et~al.}(2008)\citenamefont
  {Campbell}, \citenamefont {Groenenboom}, \citenamefont {Lu}, \citenamefont
  {Tsikata},\ and\ \citenamefont {Doyle}}]{08CaGrLu}%
  \BibitemOpen
  \bibfield  {author} {\bibinfo {author} {\bibfnamefont {W.~C.}\ \bibnamefont
  {Campbell}}, \bibinfo {author} {\bibfnamefont {G.~C.}\ \bibnamefont
  {Groenenboom}}, \bibinfo {author} {\bibfnamefont {H.-I.}\ \bibnamefont {Lu}},
  \bibinfo {author} {\bibfnamefont {E.}~\bibnamefont {Tsikata}}, \ and\
  \bibinfo {author} {\bibfnamefont {J.~M.}\ \bibnamefont {Doyle}},\ }\href
  {\doibase 10.1103/PhysRevLett.100.083003} {\bibfield  {journal} {\bibinfo
  {journal} {Phys. Rev. Lett.}\ }\textbf {\bibinfo {volume} {100}},\ \bibinfo
  {pages} {083003} (\bibinfo {year} {2008})}\BibitemShut {NoStop}%
\bibitem [{\citenamefont {Kozyryev}\ \emph {et~al.}(2015)\citenamefont
  {Kozyryev}, \citenamefont {Baum}, \citenamefont {Matsuda}, \citenamefont
  {Olson}, \citenamefont {Hemmerling},\ and\ \citenamefont {Doyle}}]{15KoBaMa}%
  \BibitemOpen
  \bibfield  {author} {\bibinfo {author} {\bibfnamefont {I.}~\bibnamefont
  {Kozyryev}}, \bibinfo {author} {\bibfnamefont {L.}~\bibnamefont {Baum}},
  \bibinfo {author} {\bibfnamefont {K.}~\bibnamefont {Matsuda}}, \bibinfo
  {author} {\bibfnamefont {P.}~\bibnamefont {Olson}}, \bibinfo {author}
  {\bibfnamefont {B.}~\bibnamefont {Hemmerling}}, \ and\ \bibinfo {author}
  {\bibfnamefont {J.~M.}\ \bibnamefont {Doyle}},\ }\href {\doibase
  10.1088/1367-2630/17/4/045003} {\bibfield  {journal} {\bibinfo  {journal}
  {New J. Phys.}\ }\textbf {\bibinfo {volume} {17}},\ \bibinfo {pages} {045003}
  (\bibinfo {year} {2015})}\BibitemShut {NoStop}%
\bibitem [{\citenamefont {Rellergent}\ \emph {et~al.}(2013)\citenamefont
  {Rellergent}, \citenamefont {S}, \citenamefont {Schowalter}, \citenamefont
  {Kotochigova}, \citenamefont {Chen},\ and\ \citenamefont
  {Hudson}}]{13ReSuSc}%
  \BibitemOpen
  \bibfield  {author} {\bibinfo {author} {\bibfnamefont {W.}~\bibnamefont
  {Rellergent}}, \bibinfo {author} {\bibfnamefont {S.}~\bibnamefont {S}},
  \bibinfo {author} {\bibfnamefont {S.}~\bibnamefont {Schowalter}}, \bibinfo
  {author} {\bibfnamefont {S.}~\bibnamefont {Kotochigova}}, \bibinfo {author}
  {\bibfnamefont {K.}~\bibnamefont {Chen}}, \ and\ \bibinfo {author}
  {\bibfnamefont {E.~R.}\ \bibnamefont {Hudson}},\ }\href {\doibase
  10.1038/nature11937} {\bibfield  {journal} {\bibinfo  {journal} {Nature}\
  }\textbf {\bibinfo {volume} {495}},\ \bibinfo {pages} {490} (\bibinfo {year}
  {2013})}\BibitemShut {NoStop}%
\end{thebibliography}%

\end{document}